\pdfoutput=1
\documentclass[aps,prl,reprint,amsmath,amssymb,superscriptaddress]{revtex4-1}
\usepackage{bm}
\usepackage{graphicx}
\usepackage[colorlinks]{hyperref}
\usepackage{mathrsfs}
\usepackage{times}
\usepackage[dvipsnames]{xcolor}
\usepackage{float}

\newcommand{\dd}{{\rm d}}

\newcommand{\hz}[1]{{\color{red} \textbf{[HZ: #1]}}}

\newcommand{\be}{\begin{equation}}
\newcommand{\ee}{\end{equation}}

\usepackage{physics}

\graphicspath{{Figures/}}

\begin{document}

\newcommand{\titleinfo}{Confinement induced impurity states in spin chains}

\title{\titleinfo}
\author{Joseph Vovrosh}
\affiliation{Blackett Laboratory, Imperial College London, London SW7 2AZ, United Kingdom}

\author{Hongzheng Zhao}
\affiliation{Blackett Laboratory, Imperial College London, London SW7 2AZ, United Kingdom}

\author{Johannes Knolle}
\affiliation{Department of Physics TQM, Technical University of Munich, 85748 Garching, Germany}
\affiliation{Munich Center for Quantum Science and Technology (MCQST), Schellingstr. 4, D-80799 M{\"u}nchen, Germany}
\affiliation{Blackett Laboratory, Imperial College London, London SW7 2AZ, United Kingdom}

\author{Alvise Bastianello}
\affiliation{Department of Physics and Institute for Advanced Study, Technical University of Munich, 85748 Garching, Germany}
\affiliation{Munich Center for Quantum Science and Technology (MCQST), Schellingstr. 4, D-80799 M{\"u}nchen, Germany}

\begin{abstract}
Quantum simulators hold the promise of probing central questions of high-energy physics in tunable condensed matter platforms, for instance the physics of confinement. Local defects can be an obstacle in these setups harming their simulation capabilities. However, defects in the form of impurities can also be useful as  probes of  many-body correlations and may lead to fascinating new phenomena themselves.
Here, we investigate the interplay between impurity and confinement physics in a basic spin chain setup, showing the emergence of exotic excitations as impurity-meson bound states with a long lifetime.
For weak confinement, semiclassical approximations can describe the capture process in a meson-impurity scattering event. In the strong-confining regime, intrinsic quantum effects are visible through the quantization of the emergent bound state energies which can be readily probed in quantum simulators.
\end{abstract}

\maketitle

%%%%%%%%%%%%%%%%%%%%%%%%%%%%%%%%%
%%%%%%%%%%%%%%%%%%%%%%%%%%%%%%%%%

\paragraph{Introduction---}

The advent of experimental platforms with high precision control has lead to remarkable progress on the path towards faithful quantum simulators~\cite{feynman2018simulating,Georgescu2014quantum}. Current quantum devices, in principle, have enough qubits to access real time quantum many-body dynamics beyond the reach of classical devices~\cite{Altman2021quantum}. However, identifying problems which are of practical importance and whose simulation is feasible with current technology remains a challenge. A growing interest is directed towards realizing prototypical examples of high-energy physics in quantum simulators~\cite{jordan2012quantum,banuls2020simulating,yang2020observation}. The hope is that experimental advances will overcome the limitations in system sizes and the need for simplified toy models for eventually simulating the fascinating, but extremely challenging, physics of strongly coupled gauge theories~\cite{brambilla2014qcd} probed at large hadron colliders.

Of particular interest is the phenomenon of confinement: the interaction strength between quarks grows as a function of their separation and, thus, they cannot be observed in isolation but only as constituents of baryons or mesons. A simplified version sharing key characteristics of confinement also appears in condensed matter physics \cite{PhysRevD.18.1259,DELFINO1996469,DELFINO1998675}, for example as domain-wall confinement in spin chains~\cite{fonseca2003ising,rutkevich2008energy}. Signatures of the ensuing meson bound states have been famously observed in inelastic neutron scattering experiments~\cite{coldea2010quantum,lake2010confinement}. 
Admittedly, these one-dimensional systems are a crude oversimplification of true hadronic physics, they nevertheless possess the same basic ingredients and present an ideal testbed for available quantum simulators, e.g. for probing real time signatures of confinement~\cite{kormos2017real}. Indeed, new quench protocols have enabled recent quantum simulations of confinement in trapped ions~\cite{martinez2016real,tan2021domain} as well as in superconducting platforms~\cite{vovrosh2021confinement}. 
The recent research program focusing on real time dynamical aspects of confinement has in itself lead to new discoveries, e.g. novel non-equilibrium phenomena with anomalously slow information spreading~\cite{kormos2017real,james2019nonthermal,liu2019confined,mazza2019suppression,PhysRevB.102.014308, PhysRevB.102.041118,viti21}, false vacuum decay \cite{Sinha_2021,lagnese2021false,milsted2021collisions,rigobello2021entanglement,Javier_Valencia_Tortora_2020,pomponio2021bloch} and dynamical phase transitions \cite{PhysRevResearch.2.033111, PhysRevB.97.174401, hashizume2020dynamical}.

However, most of these studies focused essentially on single-meson physics and the interplay among mesons themselves, or with other constituents, is yet to be fully addressed.
These are crucial questions from the perspective of simulating high-energy experiments which are based on scattering events. In a broader context, genuine many-body physics of confined excitations can rightfully expected to be far richer -- and challenging -- than the already intriguing single-meson phenomena.
Very recently, this program has been started in Refs. \cite{surace2021scattering,karpov2020spatiotemporal,milsted2020collisions} with the investigation of mesonic scattering in the Ising chain with a tilted magnetic field.

There, domain walls are pairwise confined forming the analogue of two-quark mesons and the quantized internal degrees of freedom label different mesonic species.
Due to the composite nature of mesons, the scattering of two wavepackets is deeply inelastic with the possibility of exciting mesonic internal degrees of freedom different from the injected ones.
Nevertheless, only asymptotic states of two-quark mesons can be obtained and particles formed by a larger number of constituents, i.e. baryon analogues, do not exist in the spectrum of the theory (see however Refs. \cite{liu2020realizing,Rutkevich_2015}).

In condensed matter, scattering processes with the possibility of bound state formation appear prominently in the context of impurities. On a practical side, any experimental setup unavoidably features defects whose effect must be understood. More interestingly in the quantum simulation realm, impurities in the form of localized potentials can themselves serve as \textit{probes} of the quantum properties of the host. Famous examples are the distinct impurity response of singlet versus triplet superconductors (SCs)~\cite{anderson1959theory,mackenzie1998extremely}, the characteristic impurity signal establishing the d-wave symmetry of high-temperature cuprate SCs~\cite{pan2000imaging}, as well as the local response of fractionalized edge spins in Haldane spin chain compounds~\cite{hagiwara1990observation}. Alternatively, impurities as truly dynamical objects strongly couple to the background matter, giving rise to the venerable Kondo effect~\cite{kondo1964resistance} or polaronic bounds states for mobile but heavy impurities~\cite{landau1948effective,schmidt2018universal}.  

\begin{figure}[t]
\includegraphics[width=0.8\linewidth]{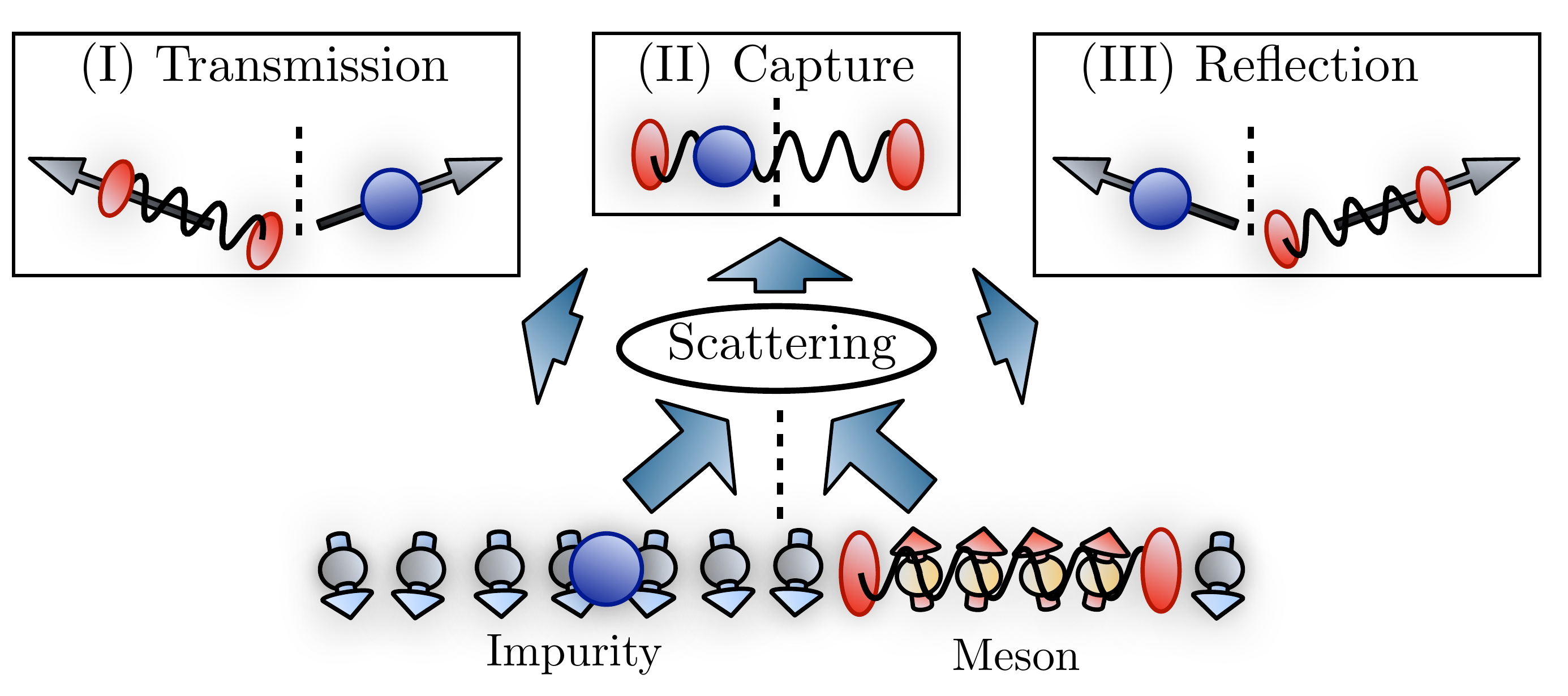}
	\caption{Pictorial representation of the impurity-meson scattering and bound state formation. Besides the instantaneous transmission (I) and reflection (III), the colliding impurity can quantum-tunnel through the first quark and form a metastable bound state (II). This composite excitation decays when the impurity tunnels again through one of the two quarks.
	}
	\label{fig_cartoon}
\end{figure}

Here, we study the interplay of confinement and impurity dynamics in a spin chain set-up. Similarly to the direct meson-meson interaction, scattering of a meson with an impurity is deeply inelastic. Apart from the main transmission (I) and reflection (III) processes of a scattering event, see Fig.\ref{fig_cartoon},  we show that in our system a nontrivial capture  (II) may appear because the different nature of the impurity and the confined excitations allows for the creation of new composite particles, with a long lifetime. Thus, the confinement-induced impurity bound state -- an elementary example of baryon formation -- can serve as a new probe of confinement physics, which is readily implementable in available quantum simulators.

\paragraph{The model and impurity set-up---}
As a paradigmatic model of confinement dynamics in a spin chain setting, we focus on the Ising chain in transverse and longitudinal magnetic fields
\be\label{eq_ising}
H_\text{Ising}=-\sum_j \sigma_j^z\sigma_{j+1}^z - h_z\sum_j \sigma_j^z - h_x\sum_j \sigma_j^x.
\ee
In the absence of a longitudinal field $h_z=0$, the model is equivalent to non-interacting fermions and features a phase transition at $h_x=1$. Above the critical point $h_x >1$, the
fermions describe magnonic excitations which become domain
walls when considering $h_x < 1$. Domain walls, or kinks (we use both names interchangeably), are topological excitations interpolating between the two degenerate ground states: the degeneracy is weakly broken by turning on a small longitudinal field $h_z$. As such, regions of spins pointing in the wrong direction pay an energy proportional to their size and induce a linear attractive potential between kinks.
The original fermions are now pairwise confined in the natural excitations of the theory, which are readily interpreted as mesons.

This qualitative picture is surprisingly robust for finite longitudinal fields: indeed, the Hamiltonian \eqref{eq_ising} does not only add interactions to the $h_z=0$ fermionic excitations, but strictly speaking it also spoils number conservation. 
Fermion conservation at weak $h_z$ can be recovered after a non-trivial rotation of the computational basis, that order by order in perturbation theory prevents fermion production, resulting in an exponentially long mesonic lifetime \cite{PhysRevB.102.041118}.
While the form of the Hamiltonian \eqref{eq_ising} governing the bulk dynamics is crucial for the capture process, the precise form of the impurity is not, as clarified by the forthcoming semiclassical picture. Nevertheless, for the sake of concreteness we focus on a simple local spinflip coupling of the impurity to the spin chain,
\be 
V=\sum_j (h_x-d)\sigma_j^x c^\dagger_j c_j \, ;\,\,\, H_\text{I}=-\sum_j \tau (c^\dagger_{j+1}c_j+c^\dagger_j
c_{j+1}).
\ee
and the basic tight-binding Hamiltonian allows the study of a dynamical impurity
with hopping strength $\tau$.
The entire system evolves with $H=H_\text{Ising}+H_\text{I}+V$. For $d=0$, the spin flip is entirely suppressed at the impurity's position.
For a single impurity, $c_j$ can be equivalently chosen to obey standard fermionic or bosonic commutation relations.
\begin{figure}[b!]
	\includegraphics[width=0.99\linewidth]{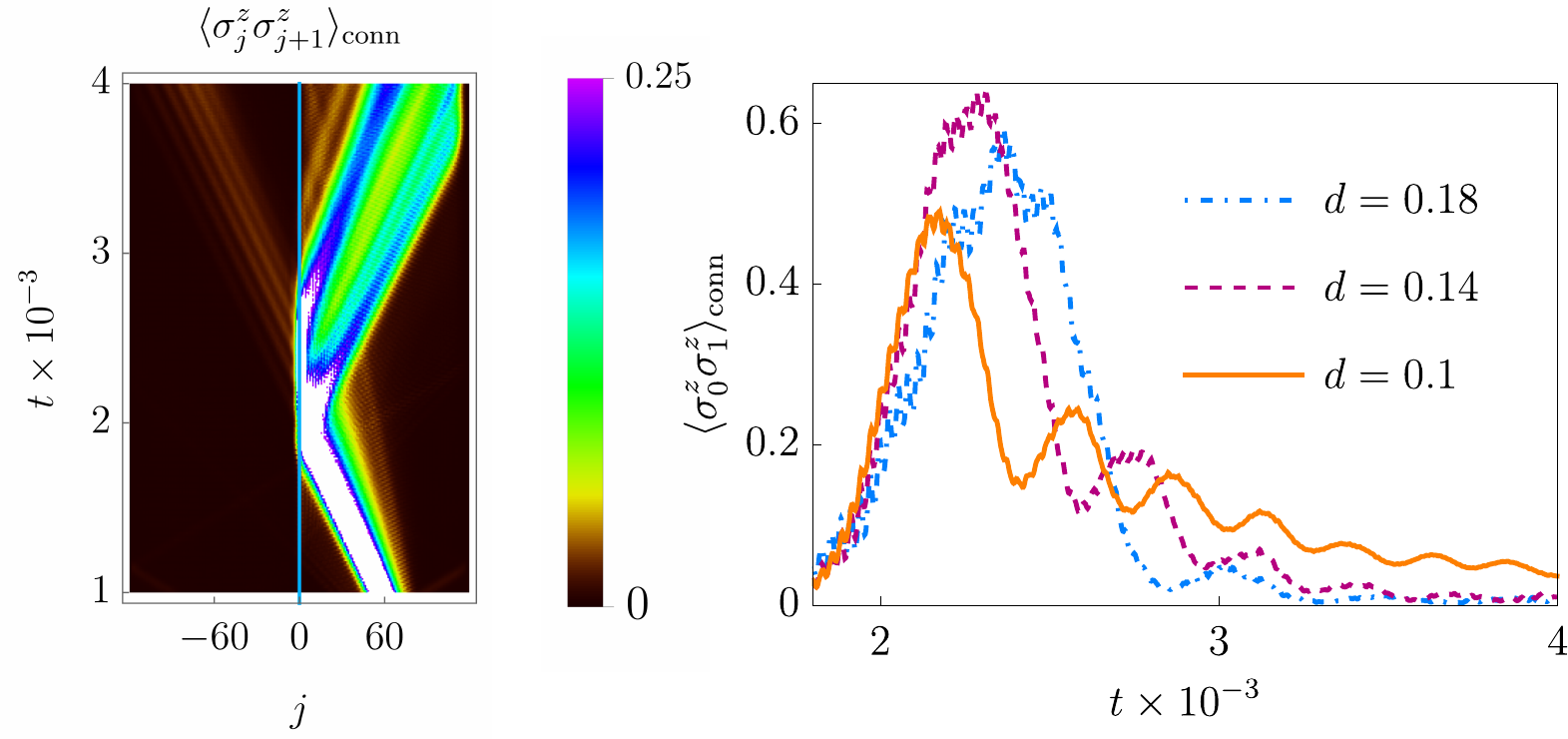}
	\caption{TEBD simulation of the scattering of a meson against an infinitely massive impurity. The correlator $\langle \sigma_j^z\sigma_{j+1}^z\rangle$ (connected part) tracks the position of kinks.
	We chose $h_x=0.3,\,h_z=0.12.$ and different values of the defect strength.
	Left: density plot for $d=0.1$, showing kinks are trapped on the impurity at the origin for very long times.
	Right: kink density on the defect for different defect strengths. After a large peak corresponding to the impact of the meson wavepacket with the defect, some signal remains trapped for longer times. As the defect's strength is increased, the magnitude of the trapped signal is reduced, but its lifetime is increased. Details on the wavepacket implementation can be found in the SM \cite{Note1}.}
	
	\label{fig_tensor_networks}
\end{figure}
\paragraph{Confinement and metastable trapping---}
In a scattering event, the meson can be transmitted through the impurity if both domain walls tunnel through it. However, if the kink-impurity transmission probability $T$ is small, the simultaneous tunneling of both kinks is further suppressed $\sim T^2$, see Fig.~\ref{fig_cartoon}.
In the absence of confinement, the transmitted and reflected fermions will eventually leave the scattering region, but a confining force in combination with a small transmission rate can trap the two kinks on opposite sides of the impurity for very long times.
To substantiate the intuitive picture, we first consider the limit of an infinitely massive impurity $\tau=0$ where the defect loses any dynamics and remains pinned. We start by numerically simulating the meson-impurity scattering with the time evolving block decimation (TEBD) method \cite{vidal2004efficient,hauschild2018efficient}. 

Details on the numerical implementation and the wavepacket preparation are given in the Supplementary Material (SM) \footnote{Supplementary Material at [url] for two-kink subspace treatment; Truncated Wigner approach; details on the numerical methods}.
After shooting a meson at the impurity (see Fig. \ref{fig_tensor_networks}) we see that the part of the wavepacket can be trapped within the defect region for long times, depending on the confining force and the defect strength. We experience that for stronger defects the signal is mostly reflected, but a small part of it remains trapped for longer times, suggesting the formation of the sought metastable bound state.
Nonetheless, it is hard to properly control the wavepacket initialization with tensor networks and explore very large time scales.
This further motivates us in quantitatively investigating the trapping mechanism with analytical means and seeking for further numerical evidence within the few-domain walls approximation.

\paragraph{The semiclassical approach---}
In our spin chain the classical limit is approached for vanishing longitudinal field.
In the absence of confinement, the excitations are free fermions with dispersion law $\epsilon(k)=2\sqrt{(h_x-\cos k)^2+\sin^2 k}$, which is then promoted to a classical kinetic energy. For $h_z\ne 0$, one can treat the fermions as point-like classical particles governed by the Hamiltonian  \be
\mathcal{H}_\text{cl}=\epsilon(k_1)+\epsilon(k_2)+\chi|j_1-j_2|,
\ee
with $\chi=2h_z \bar{\sigma}^z$ and $\bar{\sigma}^z=(1-h_x^2)^{1/8}$ being the expectation value of $\langle \sigma^z\rangle$ in the symmetry broken phase at $h_z=0$~\cite{rutkevich2008energy}. Notice that in the classical case the coordinates $j_1,j_2$ are now continuous.
Short-range corrections to the interactions are present \cite{rutkevich2008energy}, but they can be neglected in the semiclassical limit.
We treat the static impurity as a point-like scattering center, which transmits a fermion with probability $T(k)$. Aside from the randomness in the scattering process, the fermions are evolved with the deterministic classical equation of motion.
\begin{figure}[t]
	\includegraphics[width=1\columnwidth]{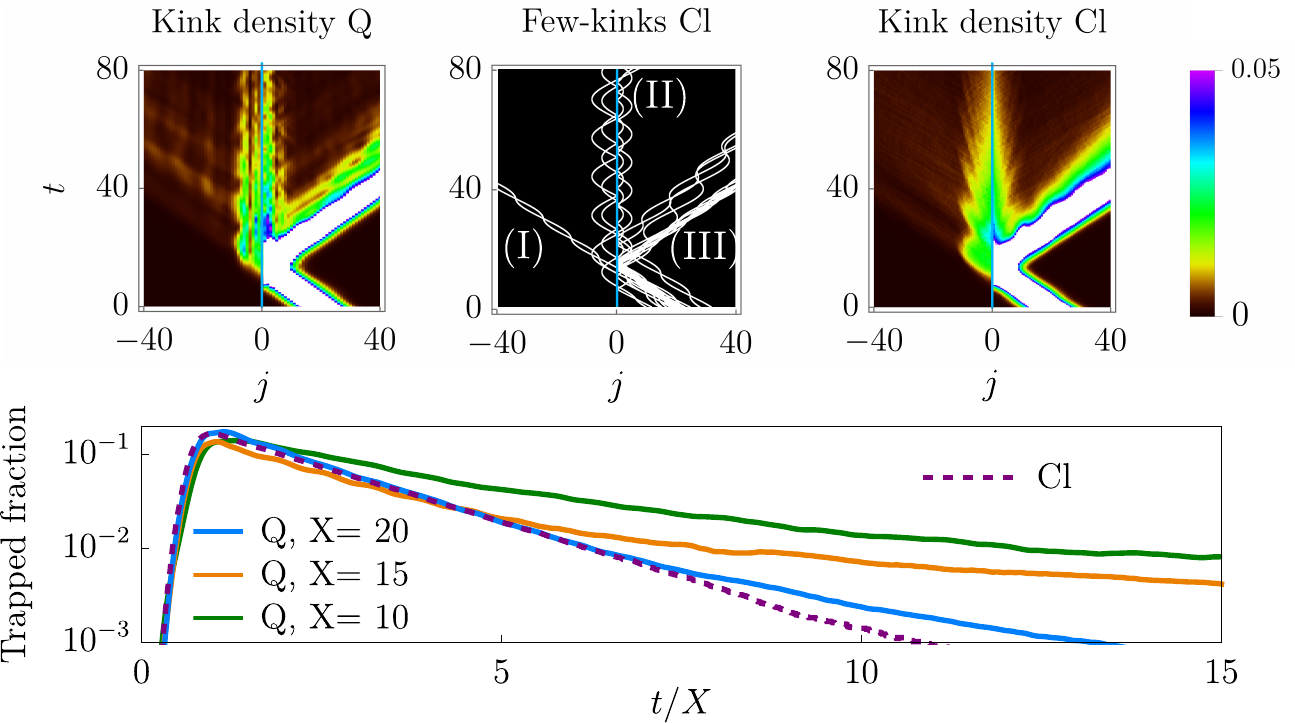}
	\caption{ 
In the two kink subspace, we initialize the two fermions in a wavepacket centered at $X$ and shoot it at a static impurity ($\tau=0$). $X$ is used as control parameter to attain the classical limit $X\to \infty$, by increasing the variance of the wavepacket and reducing the confining potential $\chi \propto X^{-1}$, see SM \cite{Note1} for details.
Top: the scattering event at short times is analyzed by plotting the fermion density for the largest $X=20$. In particular: \emph{left} quantum simulation, \emph{middle} domain wall trajectories for few classical events highlighting the three scattering process sketched in Fig.\ref{fig_cartoon}.  While most of mesons are reflected (III), some are transmitted (I) and others remain trapped (II). \emph{Right:} saturation to a large number of classical events ($\sim 4\times 10^4$).
Bottom: trapped fraction $\sum_{j_1\le 0,j_2>0}|\psi(j_1,j_2)|^2$ as a function of time. We show the quantum curves for $X=10,15,20$ and the classical curve for $X=20$, since only small differences were observed for the other choices.
The time is measured in units of $h_x^{-1}$.}
	\label{fig_twokink}
\end{figure}
We stress that $T(k)$ is not the mesonic transmission rate, but the much simpler one-particle tunneling computed in the absence of confinement.
We now consider the lifetime of an already trapped meson, which is most conveniently labeled by the momenta of the two fermions at the moment of impact with the impurity $(k,q)$. 
Let $P_{t=0}(k,q)$ be the probability of forming the boundstate immediately after the scattering, we are now interested in addressing the probability that it remains bound after a time $t$.
Notice that, in the case of a static impurity $\tau=0$, the fermions scatter with the impurity always with momenta $(k,q)$ for the whole bound state lifetime.
Bound states with the longest lifetime are characterized by small transmission probability of the fermions. Their lifetime can thus be computed as the probability that neither of the two fermions is transmitted.
Hence, the probability of being trapped at time $t$ is (see SM \cite{Note1} for details)
\be\label{eq_M_lifetime}
P_t(k,q)=\exp\left[-t\frac{\chi}{2}\big(|k|^{-1} T(k)+|q|^{-1}T(q)\big)\right]P_{t=0}(k,q)\, .
\ee
At low momentum, the discrete fermion-impurity Hamiltonian can be treated in the continuum $H_\text{eff}=-(2m_\text{eff})^{-1}\partial_x^2+c_\text{eff}\delta(x)$, leading to a quadratically vanishing transmission $T(k)\propto k^2$ as $k\to 0$.
Therefore, small momenta have a divergent lifetime.
Following this argument, one could expect a power law decay of the total trapped fraction $\int \dd k \dd q P_t(k,q)$, but this is not the case because the two momenta $(k,q)$ are not independent.
In particular, it is not possible to have $k\simeq q \simeq 0$ at the same time and the exponential decay is restored (see SM \cite{Note1} for details).
To determine the initial trapped probability $P_{t=0}$, the full time evolution of the colliding meson must be addressed. Remarkably, $P_{t=0}(k,q)$ can be explicitly computed in the limit of broad wavapackets~\cite{Note1}.

\paragraph{The two-kink subspace---}
To test the general semiclassical treatment and clarify the nature of quantum corrections, we revert to the weak transverse field regime where the Ising dynamics can be projected onto the few-kink subspace~\cite{rutkevich2008energy}.
For weak $h_x$, the fermionic excitations are well approximated by domain walls, hence we focus on the states $\ket{j_1,j_2} = \ket{\uparrow...\uparrow\downarrow_{j_1}...\downarrow\uparrow_{j_2}...\uparrow}$ and the wavefunction $\psi(j_1,j_2)=\langle j_1,j_2|\psi\rangle$.
By projecting the Ising Hamiltonian in this subspace, one finds $\langle j_1,j_2|H_\text{Ising}|j_1,j_2\rangle=\chi|j_1-j_2|+\text{const.}$ and $\langle j_1\pm 1,j_2|H_\text{Ising}|j_1,j_2\rangle=\langle j_1,j_2\pm 1|H_\text{Ising}|j_1,j_2\rangle=-h_x$. Likewise, the hopping amplitude on the defect is obtained by replacing $h_x\to d$.
There are several advantages within this approximation. First, the effective two-body problem can be simulated for large system sizes and for long times. Second, one has far better control of the form of the initial wavepacket. Third, the fermion-impurity transmission probability $T(k)$ can be exactly computed.
 \begin{figure}[t]
 \begin{centering}
	\includegraphics[width=1\linewidth]{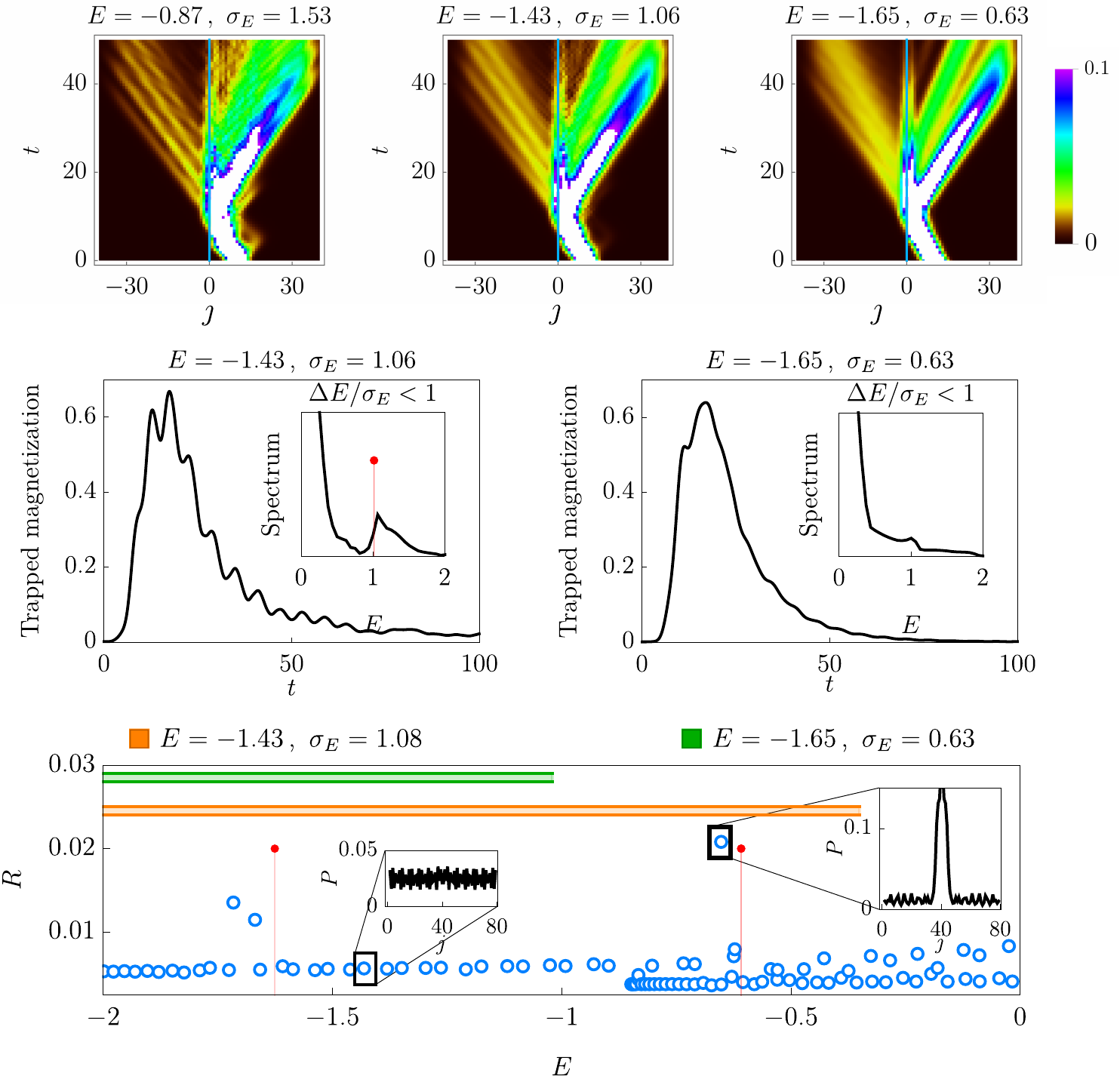}
	\end{centering}
	\caption{ 
	\label{fig_feathers}
   In the two kink subspace, we fix $\chi/h_x=0.5$ and $d/h_x=0.4$. Top: wavepackets of similar envelope (see SM \cite{Note1} for details), but different mean energy $E$ and variance $\sigma_E$ are shot at the defect. The density plot of the fermion density is shown.
    Middle: for the middle and rightmost parameter choice displayed in the density plots, we show the trapped magnetization $\sum_{j_1\le 0,j_2>0}|j_1-j_2||\psi(j_1,j_2)|^2$ as a function of time. Insets: Fourier transform (vertical axis constant but arbitrary units), the pins track the energy differences of the semiclassical quantized bound state energies within one sigma from the wavepacket mean energy. As $\sigma_E$ is reduced, fewer metastable states are excited and the oscillations due to interferences are damped. The time is measure in units of $h_x^{-1}$.
    Bottom: on a system of $L=80$ sites with periodic boundary conditions (defect at site $40$), we numerically compute the participation ratio $R=\sum_{j_1<j_2}|\psi(j_1,j_2)|^4$ of the energy eigenstates. Pins are the semiclassical quantized energies. Bound states have large $R$ when compared with naive asymptotic states. The small oscillating tails of the density profile leaving the defect region (inset, $P(j)=\sum_{j_1\le j}|\psi(j_1,j)|^2+\sum_{j_2>j}|\psi(j,j_2)|^2$) lead to the finite lifetime. Green and orange bar shown the extension of the energy interval within one sigma from the average energy probed by the two plots in the middle. 
	}
\end{figure}
In SM~\cite{Note1} we show how a standard truncated Wigner approximation~\cite{POLKOVNIKOV20101790} allows to quantitatively connect the initial quantum mechanical wavepacket with the proper classical phase space distribution.
In Fig.~\ref{fig_twokink} we provide the quantum-classical comparison, showing good agreement.

\paragraph{Beyond semiclassics and bound state requantization---}
In Fig. \ref{fig_feathers} we consider a scattering event in the two kink subspace, but with larger longitudinal field and far from the semiclassical regime. Pronounced oscillations appear in the density of fermions leaving the defect and in the time evolution of the trapped magnetization. We will now show that these frequencies are in good qualitative agreement with a semiclassical quantization of the bound state energies.
From the classical perspective, a trapped meson is a pair of fermions on opposite sides of the impurity, see Fig. \ref{fig_cartoon} (II).
Each of the two fermions feels a constant force pulling towards the barrier, which acts as a hard wall until a tunneling event takes place. 
Pushing this interpretation to the quantum regime, we can write the single particle time-independent Schrodinger equation for the right fermion as $E_{n}\psi(j)=-h_x(\psi(j+1)+\psi(j-1))+\chi |j|\psi(j)$, valid for $j>0$ and with boundary condition $\psi(0)=0$. An analogue equation holds for the left fermion.
This equation can be analytically solved for the quantized energies $\{E_n\}_{n=0}^\infty$ \cite{Note1}. Therefore, the energy of a metastable state is labelled by two quantum numbers
$E_{n_1,n_2}=E_{n_1}+E_{n_2}-\chi$, with the $\chi-$shift taking into account the string tension on the defect link.
In Fig. \ref{fig_feathers}, we show that the oscillation frequencies are close to the energy differences of the quantized metastable states.
We selectively excite the quantized metastable states by shooting wavepackets with mean energy $ E$ and narrow variance $\sigma_E$. If $\sigma_E$ is large enough to excite more than one metastable energy, clear oscillations are produced. See also SM \cite{Note1} for further analysis.
\begin{figure}[t]
	\includegraphics[width=1\linewidth]{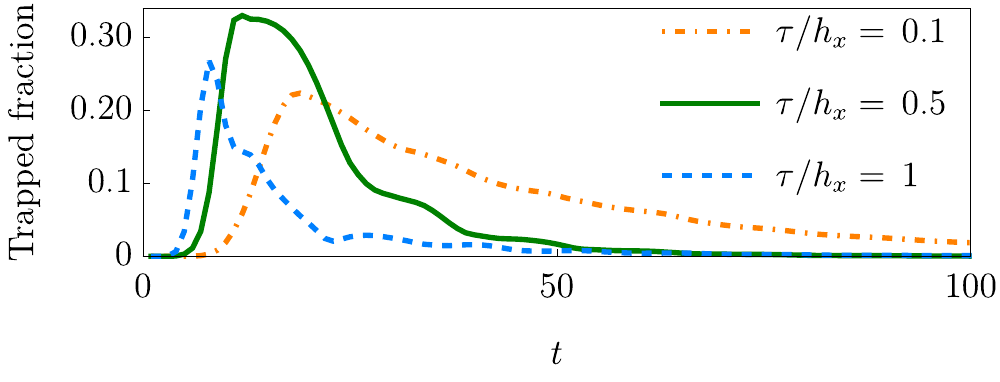}
	\caption{ \label{fig_movimp}
	In the total zero momentum sector, we consider the scattering of a mobile impurity with a meson. The initial conditions, defect strength and confinement are kept fixed while varying $\tau$ and are such that for $\tau=0$ they match Fig. \ref{fig_twokink} with $X=20$. Further details in SM \cite{Note1}.
	}
\end{figure}
\paragraph{The dynamical impurity case---}
We finally address the case of a truly dynamical impurity with $\tau\ne 0$.
For the sake of simplicity, we consider the weak transverse field regime and the subspace of a fixed constant momentum, which can be  addressed in the  two-kink approximation (see SM \cite{Note1}). 
The semiclassical picture we discussed for a static impurity can be readily applied to the case $\tau\ne 0$ by using the equation of motion and the fermion-impurity transmissivity rate, which can be exactly computed for weak $h_x$ \cite{Note1}.
As expected, $\tau\ne 0$ increases the relative impurity-fermion motility, increasing the transmission rate and shortening the lifetime of the metastable state \eqref{eq_M_lifetime}, which is nevertheless still present, see Fig. \ref{fig_movimp}.
However, the simple result of Eq. \eqref{eq_M_lifetime} cannot be applied any longer, since each fermion is not simply reflected as $k\to -k$ due to the momentum exchange with the impurity. In contrast, the fermion-impurity scattering process is largely affected by the defect's velocity \cite{PhysRevLett.120.060602,PhysRevB.98.064304} and, outside the two-kink approximation, the moving impurity can act as a moving source of excitations \cite{PhysRevB.101.085139}.

\paragraph{Conclusions and outlook---}
We demonstrated the creation of exotic and long-lived composite particles from the interplay of confinement and impurity dynamics. We quantitatively framed the problem within a simple semiclassical picture and describe quantum corrections by semiclassically quantizing the metastable bound state eigenenergies. 

Our predictions should be readily observable in state-of-the-art quantum simulators, where defects can be easily engineered~\cite{martinez2016real,Bernien2017}, and which opens the path for addressing many interesting questions.
First, the effect of a truly moving impurity has only been partially investigated and could lead to interesting phenomena arising from the interplay between the defect's velocity and the lightcone of the mesons~\cite{PhysRevLett.120.060602,PhysRevB.98.064304,PhysRevB.101.085139}. Besides, the energy exchange between the impurity and the chain can alter the Schwinger effect \cite{PhysRev.82.664} (see also \cite{Sinha_2021,lagnese2021false,milsted2021collisions,rigobello2021entanglement,Javier_Valencia_Tortora_2020,pomponio2021bloch}) by converting the impurity kinetic energy in a quark-antiquark pair on the false vacuum  and spur thermalization.
Another interesting scenario to be investigated is whether the impurity itself can be confined with a companion: when two mesons of different particle species scatter, each quark can act as the impurity for the other species, exciting four-quark (or more) mesonic particles. Finally, it would be worthwhile to investigate confinement induced impurity states beyond the simple the simple spin chain context, e.g. in quantum chromodynamics. 
The physics of impurities was famously dismissed as 'squalid state physics' from a high-energy perspective~\cite{johnson2001new}, but in the context of quantum simulations thereof it might turn out to be exceptionally useful. 
%%%%%%%%%%%%%%%%%%%%%%%%%%%%%%%%%

\paragraph{Acknowledgements---}  
 JV acknowledges the Samsung Advanced Institute of Technology Global Research Partnership and travel support via the Imperial-TUM flagship partnership. AB acknowledges support from the Deutsche Forschungsgemeinschaft (DFG, German Research Foundation) under Germany's Excellence Strategy-EXC-2111-390814868. HZ acknowledges support from a Doctoral-Program Fellowship of the German Academic Exchange Service (DAAD). Tensor network calculations were performed using the TeNPy Library \cite{hauschild2018efficient}.
%%%%%%%%%%%%%%%%%%%%%%%%%%%%%%%%%

\bibliography{biblio}

%%%%%%%%%%%%%%%%%%%%%%%%%%%%%%%%%

\onecolumngrid
\newpage

\setcounter{equation}{0}            % reset equation counter
\setcounter{section}{0}    % reset section counter
\renewcommand\thesection{\arabic{section}}    % puts letters as section numbering
\renewcommand\thesubsection{\arabic{subsection}}    % puts letters as section numbering
\renewcommand{\thetable}{S\arabic{table}}
\renewcommand{\theequation}{S\arabic{equation}}
\renewcommand{\thefigure}{S\arabic{figure}}
\setcounter{secnumdepth}{2}  % if the subsections need to be numbered

\begin{center}
{\large Supplementary Material \\ 
\titleinfo} \\
Joseph Vovrosh, Hongzheng Zhao, Johannes Knolle, Alvise Bastianello
\end{center}

\bigskip
\bigskip
\bigskip

%%%%%%%%%%%%%%%%%%%%%%%%%%%%%%%%%

In this Supplementary Material we discuss in detail the numerical and analytical methods used in our work. In particular, it is organized in the following way:

\textbf{Section I:} Details on the two-domain wall approximation used for simulations through out this work; Derivation of the scattering matrix for collisions between a fermion and the impurity; The semiclassical quantization of the metastable bound state energies; Extra simulations exploring the presence of metastable states. Finally, we give the details about Figs. \ref{fig_feathers} and \ref{fig_movimp} of the main text.

\textbf{Section II}: 
Details on the semiclassical approximation (Truncated Wigner approach); Capture and decay of the metastable state within the classical approximation; Further details on the quantum-classical comparison of Fig. \ref{fig_twokink}.

\textbf{Section III}: Details on the tensor network simulations of the full Hilbert space; The initialization of the wavepacket; Further details on Fig. \ref{fig_tensor_networks}.

\section{The weak transverse field limit: the two domain wall approximation}

In the limit of weak transverse field, the fermionic excitations become the domain walls \cite{kormos2017real}, as we discuss in the main text. Let us consider the subspace with only two domain walls $\ket{j_1,j_2} = \ket{\uparrow...\uparrow\downarrow_{j_1}...\downarrow\uparrow_{j_2}...\uparrow}$ and a single impurity with position $y$. We label the three particle state as $|j_1,j_2,y\rangle$ and wavefunction $\psi(j_1,j_2,y)=\langle j_1,j_2,y|\psi\rangle$. By projecting the Hamiltonian in this subspace $\hat{H}\to \hat{H}^P$ we find
\begin{multline}\label{S_twokink}
\hat{H}^P\psi(j_1,j_2,y)=-h_x \big[\psi(j_1+1,j_2,y)+\psi(j_1-1,j_2,y)+\psi(j_1,j_2+1,y)+\psi(j_1,j_2-1,y)\big]+\chi |j_1-j_2| \psi(j_1,j_2,y)+
\\
-\tau\left[\psi(j_1,j_2,y+1)+\psi(j_1,j_2,y-1)\right]+\\
-(d-h_x)\left[\delta_{j_1,y+1}\psi(j_1-1,j_2,y)+\delta_{j_1,y}\psi(j_1+1,j_2,y)+\delta_{j_2,y+1}\psi(j_1,j_2-1,y)+\delta_{j_2,y}\psi(j_1,j_2+1,y)\right].
\end{multline}
Above, the exclusion $j_1<j_2$ is enforced when needed.
The confining force is $\chi=2h_z$ and the first line describes the fermion-fermion (i.e. meson) dynamics, the second line contains the impurity dynamics and lastly the third line captures the meson-impurity interaction.
We can conveniently focus on a sector with a well defined total momentum and reduce the problem to a two dimensional one. With a slight abuse of notation we set
\be
\psi(j_1,j_2,y)\to  e^{iK(j_1+j_2+y)}\psi(j_1-y,j_2-y)
\ee

and Hamiltonian
\begin{multline}\label{S_twokinkK}
\hat{H}^P\psi(j_1,j_2)=-h_x
\big[e^{iK}\psi(j_1+1,j_2)+e^{-iK}\psi(j_1-1,j_2)+e^{iK}\psi(j_1,j_2+1)+e^{-iK}\psi(j_1,j_2-1)\big]+\chi |j_1-j_2| \psi(j_1,j_2)+\\
-\tau\big[e^{iK}\psi(j_1-1,j_2-1)+e^{-iK}\psi(j_1+1,j_2+1)\big]+\\
-(d-h_x)\big[\delta_{j_1,1}e^{-iK}\psi(j_1-1,j_2)+\delta_{j_1,0}e^{iK}\psi(j_1+1,j_2)+\delta_{j_2,1}e^{-iK}\psi(j_1,j_2-1)+\delta_{j_2,0}e^{iK}\psi(j_1,j_2+1)\big].
\end{multline}

For the sake of simplicity, in this work we always focus on the case of zero total momentum $K=0$.

\ \\ \ \\ \ \\
\textbf{The fermion-impurity scattering matrix ---} A key ingredient in the semiclassical treatment of the metastable state is the transmission amplitude of the fermion-impurity scattering. In the weak transverse field limit, this can be analytically computed.
In order to do this, we now remove the confinement in Eq. \eqref{S_twokinkK} and focus on the single kink problem $\psi(j_1,j_2)\to \psi(j_1)$.
In this case, we now look for the eigenvectors of the time-independent Schrodinger equation
\be
E\psi(j)=-(h_xe^{iK}+\tau e^{-iK})\psi(j+1)-(h_xe^{-iK}+\tau e^{iK})\psi(j-1)-(d-h_x)\left[\delta_{j,1}e^{-iK}\psi(j-1)+\delta_{j,0}e^{iK}\psi(j+1)\right].
\ee

The asymptotic scattering solution can be written as
\be
\tilde{\psi}(j)=\begin{cases}\psi_\text{in}(j)=e^{iqj}+r(q,K) e^{i\bar{q}j} \hspace{2pc} &j\le 0, \\
\psi_\text{out}(j)=t(q,K) e^{i\bar{q}j} \hspace{2pc} &j\ge 1.
\end{cases}
\ee
Notice that due to the motion of the impurity the incoming momentum $q$ is not reflected to $-q$, but rather on all the possible eigenenergy solutions $\bar{q}$
where $E=\epsilon(q)=\epsilon(\bar{q})$ and
\be
\epsilon(q)=-2h_x\cos(K+q)-2\tau \cos(K-q).
\ee
Above, we are assuming $\partial_q \epsilon(q)=v(q)>0$, the case $v(q)<0$ is symmetric under parity reflection. $r(q,K)$ and $t(q,K)$ are the reflection and transmission coefficient respectively, while $R(q,K)=|r(q,K)|^2$ and $T(q,K)=|t(q,K)|^2$ are the reflection and transmission probability. Of course, one has $R+T=1$.
The coefficients $r,t$ are determined by imposing the Schrodinger equation at $j=\{0,1\}$ and after some simple algebraic manipulations one gets the conditions

\be
\Big(\frac{(h_x e^{iK}+\tau e^{-iK})}{(d-h_x)e^{iK}}+1\Big)\psi_\text{out}(1)
=\frac{(h_x e^{iK}+\tau e^{-iK})}{(d-h_x)e^{iK}}\psi_\text{in}(1),
\ee
\be
\Big(\frac{(h_x e^{-iK}+\tau e^{iK})}{(d-h_x)e^{iK}}+1\Big)\psi_\text{in}(0)
=\frac{(h_x e^{-iK}+\tau e^{iK})}{(d-h_x)e^{-iK}}\psi_\text{out}(0)\, .
\ee

From these, the following expression for $r(q,K)$ is easily computed

\be
r(q,K)=-\frac{1-\Big(\frac{(d-h_x)}{(h_x +\tau e^{i2K})}+1\Big)^{-1}\Big(\frac{(d-h_x)}{(h_x +\tau e^{-i2K})}+1\Big)^{-1}}{1-e^{i(\bar{q}-q)}\Big(\frac{(d-h_x)}{(h_x +\tau e^{i2K})}+1\Big)^{-1}\Big(\frac{(d-h_x)}{(h_x +\tau e^{-i2K})}+1\Big)^{-1}}
\ee

and the transmission rate follows as $T(q,K)=1-|r(q,K)|^2$.
Notice that in the case of static a impurity $\tau=0$, the transmission rate becomes $K-$independent as expected and has the simple form
\be
T(q)=\frac{4\sin^2 q}{(d/h_x)^2+(d/h_x)^{-2}-2\cos(2q)}\, .
\ee
\ \\ \ \\ 
\textbf{The mesonic energies ---} The form of the mesonic energies in the bulk are well established as the solutions to the equation 
\be
\mathcal{J}_{-\nu_{k,\alpha}}(x_k) = 0
\ee
in which $\nu_{k,\alpha} = \frac{E_{k,\alpha}}{2h_x}$, $x_k = \frac{2h_x\cos\frac{k}{2}}{h_z}$ and $\mathcal{J}$ is the Bessel function of the first kind \cite{rutkevich2008energy}.

We use a similar analysis to now provide the details on the semiclassical requantization of the energies of the metastable bound states, focusing on $\tau=0$. As we discussed in the main text, we look at the metastable state as two independent fermions bouncing on a hard wall potential. In this respect, the energy of the metastable bound state is given by $E_{j_1,j_2}=E_{j_1}+E_{j_2}+\chi$ with $E_j$ the quantized energy levels of each fermion. 
Focusing on the right fermion, its wavefunction obeys the time-independent Schrodinger equation 
\be
E_j\psi_{E_j}(x)=-h_x\psi_{E_j}(x+1)-h_x\psi_{E_j}(x-1)+\chi x \psi_{E_j}(x)\hspace{2pc} x>0
\ee
and boundary condition $\psi_{E_j}(0)=0$. This equation, apart from a numerical factor, is very similar to the wavefunction in the relative coordinates of two confined domain walls, hence it can be solved in a similar manner. The eigenfunctions are parametrized by Bessel functions
\be
\psi_{E_j}(x)= J_{x-E_j/\chi}\left(-2h_x/\chi\right)\, .
\ee
The energies are then found by imposing the condition $J_{-E_j/\chi}\left(-2h_x/\chi\right)=0$.

\begin{figure}[t!]
    \centering
    \includegraphics[width=0.9\textwidth]{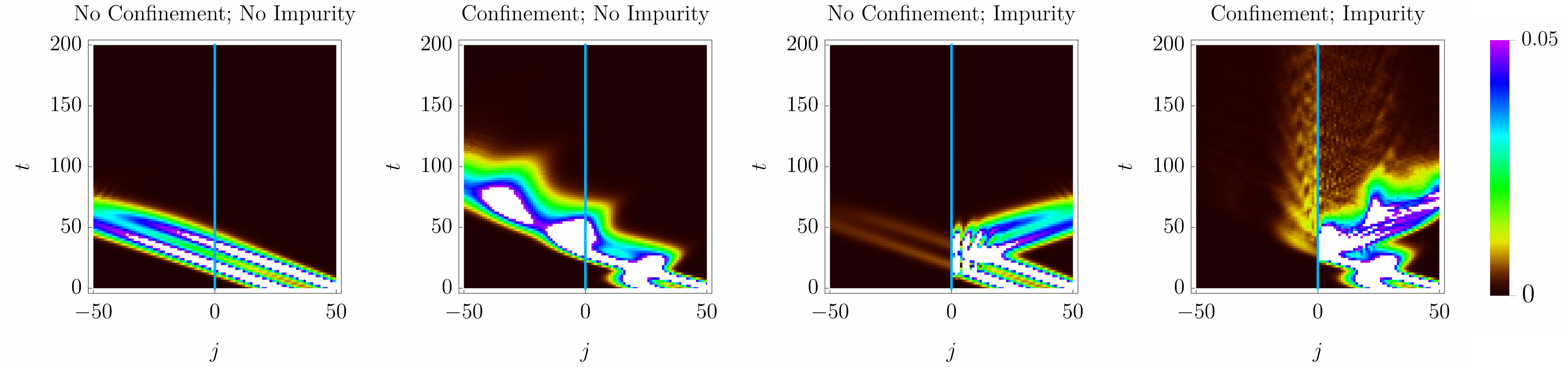}
    \caption{Two-fermion wavepacket collision against a static impurity for the same initial conditions, but in the presence and absence of confinement and defected bond (see text). Here, we focus on the domain wall probability as a function of space and time (measured in units of $h_x$).}
    \label{fig_comparison}
\end{figure}

\ \\ \ \\ \ \\
\textbf{Numerical simulations in the two-kink subspace ---} 
Within the two-kink subspace, we have great control on the numerical simulations of the meson scattering. In Fig. \ref{fig_comparison} we provide extra evidence on the possibility of forming metastable bound states through the interplay of the confinement and impurity-scattering. We initialize the two domain walls in two narrow wavepackets with momentum $-\pi/2$ and relative distance $25$ sites. The domain wall closest to the static impurity (centered at $j=0$) is placed at $25$ sites from it.
Then, we let evolve the two domain walls in four different cases. From the left to the right:  no confining force $\chi=0$ and no impurity $d=h_x$; confinement $\chi=0.1 h_x$ but no defect $d=h_x$; no confinement $\chi=0$ but non-trivial defect $d=0.2 h_x$; and finally the case where both confinement $\chi=0.1 h_x$ and defect $d=0.2h_x$ are present. In the latter case, there is clearly a finite probability of trapping the meson on the impurity.
We wish now to supplement the analysis of Fig. \ref{fig_feathers} concerning the metastable state formation in the quantum regime with further analysis.

Hence, we take the quantization of the metastable states one step further by exploring the effect on the resulting non-equilibrium dynamics of  the initial wave packet with respect to changes in its energy, $E$, as well as the variance of this energy, $\sigma_E$. In order to have optimal control we turn to a sophisticated initial wavepacket given by
\be
\bra{j_1,j_2}\ket{\psi}=\int dq\,  e^{-\sigma_k^2(q-k)^2}e^{iq(\frac{j_1+j_2}{2}-X)}\mathcal{J}_{j_2-j_1-\nu_{q,\alpha}}\bigg(\frac{2h_x}{h_z}\cos{\frac{q}{2}}\bigg)
\label{scanning_wavepacket}
\ee
where $k$ is the initial expected momentum of the wavepacket, $X$ is the initial centre of the wavepacket and $\nu_{q,\alpha} = \frac{\epsilon_{q,\alpha}}{{2h_z}}$ in which $\epsilon_{q,\alpha}$ is the energy of the two kink subspace and $\alpha$ labels the energy level \cite{vovrosh2021confinement}. This wave packet allows us to directly choose the energy level we consider but also, via the choice of $k$ and $\sigma_k$, we have good control of $E$ and $\sigma_E$.

\begin{figure}[t!]
    \centering
    \includegraphics[width=1\textwidth]{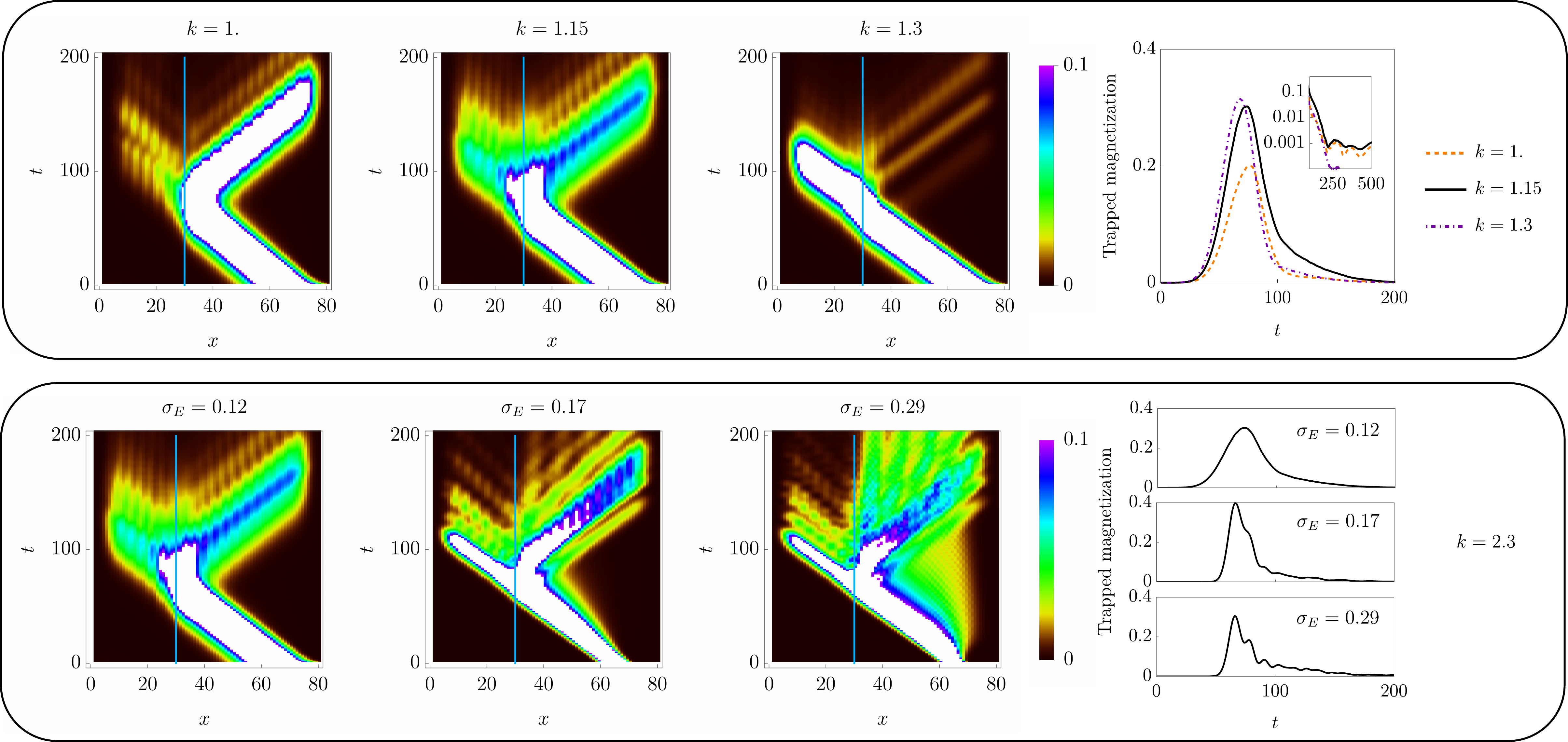}
    \caption{By initialising in the wavepacket given in Eg.\ref{scanning_wavepacket} with carefully chosen values of $k$ and $\sigma_k$ we are able to explore energy space precisely. Here we use $h_x=1$, $h_z = \frac{1}{3}$, $X = 65$ and $\alpha = 2$ with a stationary impurity located at site 30 such that $d=0.7$. In the upper panel we choose $\sigma_k^2$ such that the variance in energy space is small, $\sigma_E\sim0.12$. By varying $k$ we are able to `scan' across energy space and directly capture metastable states via a resulting long lived meson at the impurity. In the lower panel we initialise $k$ such that we observe a metastable state and slowly increase $\sigma_E$ by decreasing $\sigma_k$. This allows us to observe the resonances between different metastable states in the form of emergent oscillations in the trapped fraction of wavepacket in the non-equilibrium dynamics.}
    \label{fig_scanning}
\end{figure}

Firstly, we `scan' through energy space with a wavepacket that has a small $\sigma_E$. This directly shows that, when a wavepacket is centred on the energy of a metastable state, we observe the long lifetime of a meson trapped at the defect. As we move away form this energy, we in turn loose this signature, this can be seen in the upper panel of Fig. \ref{fig_scanning}. Furthermore, we present results of the resonances observed as we increase $\sigma_E$ of a wavepacket with energy centered such that we are capturing a metastable state. Clearly seen in lower panel of Fig. \ref{fig_scanning}, as the variance grows, more resonances are captured leading to more pronounced oscillations in the trapped fraction of the wavepacket. This is consistent with the interpretation that these oscillations are due to resonances between metastable state energies, i.e., as we increase $\sigma_E$ a large number of metastable states are excited.
\ \\ \ \\ 
\textbf{Further details about Fig. \ref{fig_feathers} and \ref{fig_movimp} ---} To generate the data for Figs. \ref{fig_feathers} and \ref{fig_movimp}, we used a wavepacket similar to Eq. \eqref{scanning_wavepacket}, but in a simpler factorized form
\be\label{S_wavepacket}
\langle j_1,j_2|\psi\rangle= e^{-\sigma^2 \left(\frac{j_1+j_2}{2}-X\right)^2}e^{i k/2 (j_1+j_2)}\phi(j_1-j_2)\, .
\ee 
This factorized wavefunction naturally arises when considering the semiclassical limit in the next section and, if the wavepacket is sufficiently smooth, it is very close to Eq. \eqref{scanning_wavepacket}.
In Fig. \ref{fig_feathers} we choose the free parameters trying to not alter the global envelope of the wavepacket, hence we kept $\sigma, X, k$ fixed and vary the energy spreading by acting on $\phi$. In particular, we chose $k=1.15$, $\sigma^2=0.08$ and $X=10$, while the defect and confinement strengths are $d/h_x=0.4$ and $\chi/h_x=2 h_z/h_x=0.5$. The time scale is in units of $h_x$, which plays the role of a global energy scale. Then, the relative wavefunction is chosen as the exact wavefunction within a finite range $\phi(|j|\le \Lambda)=\mathcal{J}_{j-\nu_{k,1}}\left(\frac{2h_x}{h_z}\cos(k/2)\right)$ and zero beyond $\phi(|j|>\Lambda)=0$. The choices $\Lambda=\{3,4,12\}$ create the wavepackets with $\sigma_E=\{1.53, 1.06,0.63\}$ respectively.

In Fig. \ref{fig_movimp} we present quantum simulations within the two-kink approximation of the metastable state lifetimes in the presence of impurities with varying mobility. We achieve this by using different values of $\tau$. Here, we use the same initial wavepacket of Fig. \ref{fig_twokink}, which we discuss in the next section. With reference to the parametrization of Eq. \eqref{S_wavepacket} and \eqref{S_phidef}, we choose $d/h_x=0.2$, $\chi/h_x=0.25$, $k=-2$, $\sigma=0.125$, $\ell=1$, $\bar{x}=3\ell$.

Numerical simulations are carried out by considering the two dimensional wavefunction on a finite segment $[-L/2,L/2]$ and the wavefunction is evolved through matrix-exponentiation of the two-kinks subspace Hamiltonian \eqref{S_twokinkK}.
In order to remove finite size corrections by simulating a true infinite system,  we add dissipation at the boundaries, thus removing the departing meson and preventing the wavepacket to return to the defect after it has been scattered away.
This approach is thoroughly discussed in the framework of tensor network simulations in Sec. \ref{S_sec_Tensor}.

\section{The semiclassical limit}

In this section we quantitatively match the semiclassical approximation against the quantum problem by means of a truncated Wigner approximation \cite{POLKOVNIKOV20101790}.
For the sake of simplicity, we focus on the two-kink subspace and consider a static impurity, but the method is readily generalized to the moving case.
Let the initial state be described by a density matrix $\hat{\rho}$, then we use a coordinate representation $|j_1,j_2\rangle$ and define the Wigner quasi-distribution as
\be\label{S_1wig}
\langle j_1+y_1/2,j_2+y_2/2|\hat{\rho}| j_1-y_1/2,j_2-y_2/2\rangle=\int \dd p_1\dd p_2\,  W(j_1,p_1,j_2,p_2) e^{i y_1 p_1+i y_2 p_2}.
\ee
In principle, one needs to enforce $j_1+y_1/2$ and all the other coordinates to be integers, but this will not be important. Indeed, classical physics emerges in the case where the wavefunction is smooth and the confinement is weak, thus coarse graining the discrete nature of the underlying lattice.
We focus on the bulk of the dynamics, leaving the impurity aside for the moment, and consider the Schrodinger equation of motion $i\partial_t \hat{\rho}=[\hat{H},\hat{\rho}]$ with $\hat{H}$ given in Eq. \eqref{S_twokink}.
By applying the equations of motion to the left hand side of Eq. \eqref{S_1wig} and expressing them in terms of the Wigner distribution, after some long but straightforward algebra one finds
\be\label{S_eqwig}
\partial_t W(j_1,p_1,j_2,p_2)+ 2 h_z\left( \sin(p_1)+ \sin(p_2)\right)W(j_1,p_1,j_2,p_2)-V'(j_1-j_2)\left(\partial_{p_1}-\partial_{p_2}\right)W(j_1,p_1,j_2,p_2)\simeq 0
\ee
with $V'(x)=\partial_x V(x)$ and $V(x)=\chi|x|$. In the derivation, one assumes $W$ to have a slow dependence on $j_1,j_2$ (smooth wavepacket) and asks $V(x)$ to be a smooth potential. In the case of confinement, $V(x)=\chi|x|$ is not smooth in the origin, but this correction vanishes in the limit of small $\chi$.
In Eq. \eqref{S_eqwig} one recognizes the classical Liouville equation for the phase space distribution $W(j_1,p_1,j_2,p_2)$ evolving with the classical Hamiltonian $\mathcal{H}_\text{cl}$ reported in the main text (in the weak transverse field regime, $\epsilon(k)=-2h_z\cos(k)$ and $v(k)=\partial_k \epsilon(k)=2h_z\sin k$).
It is useful to consider the classical equation of motion
\be
\begin{cases}\partial_t p_1= -\chi \text{sgn}(j_1-j_2)\\
\partial_t j_1=v(p_1)
\end{cases}\hspace{2pc} \begin{cases}\partial_t p_2= -\chi \text{sgn}(j_2-j_1)\\
\partial_t j_2=v(p_2)
\end{cases},
\ee
and notice the following scale invariance
\be\label{eq_S_res}
t\to X t\hspace{2pc} x_{1,2}\to X x_{1,2}\hspace{2pc} \chi \to \chi/X
\ee
where $X$ is some positive scale. Notice that this invariance holds in the classical limit, but it is broken in the quantum regime. Nevertheless, we can use it as a convenient way to attain the classical limit, by means of rescaling to larger spaces and times.
\ \\ \ \\
\textbf{The Wigner distribution of the wavepacket ---}
When comparing quantum simulations with semiclassics, it is important to correctly capture the initial conditions. Here, we provide the initial Wigner distribution for simple wavepackets that we use in the simulations. We consider pure states $\hat{\rho}=|\psi\rangle\langle \psi|$ in the factorized form already anticipated in Eq. \eqref{S_wavepacket}, but the wavefunction in the relative coordinates $\phi(x)$ is now chosen to be smooth.
For example, a convenient choice is
\be\label{S_phidef}
\phi(x)\propto\begin{cases} e^{-\frac{1}{4\ell^2} (|x|-\bar{x})^2} & \hspace{2pc} x<0 \\ 0 & \hspace{2pc} x\ge 0 \end{cases}.
\ee
Above, we ensured that the wavefunction vanishes when $j_1<j_2$.
By tuning the free parameters $\sigma,k,X,\ell,\bar{x}$ one can engineer a wavepacket of well defined momentum and control its energy.
Notice that the wavefunction is factorized in terms of the center of mass and relative coordinates, therefore the Wigner distribution has a factorized form as well
\be
W(j_1,p_1,j_2,p_2)= \mathcal{W}\left(\frac{j_1+j_2}{2},p_1+p_2\right)w\left(j_1-j_2,\frac{p_1-p_2}{2}\right)\, .
\ee
With this specific choice of wavefunction, one finds
\be
\mathcal{W}(x,p)=\frac{1}{\sqrt{2\pi \sigma^2}} \exp\left(-\frac{1}{2\sigma^2}(p-k)^2-2\sigma \left(x-X\right)^2\right).
\ee
The truncation of the Gaussian defining $\phi(x)$ in Eq. \eqref{S_phidef} prevents a simple analytical solution. However, in the limit where $e^{-\bar{x}^2/(2\ell^2)}\ll 1$ the tails can be neglected and one simply finds
\be
w(x,p)\propto \exp\left(-\frac{|x|-\bar{x})^2}{2\ell^2}-2\ell^2 p^2\right)\, .
\ee
The proportionality constant is not important and it can be fixed by ensuring the correct normalization of the state.
\ \\ \ \\
\textbf{Details on Fig. \ref{fig_twokink} ---} In Fig. \ref{fig_twokink} we compared simulations within the two-kink approximation against the truncated Wigner approach. We consider different choices of wavepackets governed by a global length scale in such a way the limit of infinitely smooth wavepacket collapses on a well defined classical limit. With reference to the wavefunction \eqref{S_wavepacket} and \eqref{S_phidef}, we chose the initial position of the wavepacket $X$ as our scaling parameter and set $d/h_x=0.2$, $\chi=5/X$, $k=-2$, $\sigma=2.5/X$, $\ell=0.05X$, $\bar{x}=3\ell$.
As we commented in Eq. \eqref{eq_S_res}, the classical equations are invariant under simultaneous rescaling of positions, time and confining strength. In the truncated Wigner, the exact invariance of the classical simulation is broken by the $X-$dependence of the momentum distribution, but it is restored in the $X\to+\infty$ limit.
\ \\ \ \\ 
\textbf{The lifetime of the metastable state ---} We now wish to provide further details about the analytical determination of the metastable state formation after a scattering event. For the sake of simplicity, we focus on the static impurity $\tau=0$, but the same analysis can be ready generalized to the moving impurity as well.
We divide the problem in two steps
\begin{enumerate}
    \item Compute the lifetime of an already trapped meson.
    \item Compute the probability of get trapped in the scattering event.
\end{enumerate}

Furthermore, we are interested only in the longest lived metastable states. Within this assumption, one can greatly simplify the analysis. If one considers an already trapped meson, it will remain trapped for long times only if the transmission probability of both fermions is very small.
Within this assumption, we can compute the two points above within these approximations
\begin{enumerate}
    \item Escape after first transmission: a trapped meson leaves the defect as soon as one of the two fermions is transmitted and it cannot be captured back.
    \item Capture after a single transmission: since transmission is unlikely, we assume the meson is captured after a single transmission event of the fermion.
\end{enumerate}

We have already discussed this simple calculation in the main text, here we quickly recap it and add some details.
It is convenient to parametrized the classical trapped meson in terms of the momenta of the fermions when they hit the barrier. Following the same notation of the main text, we call them $(k,q)$. Before transmission, each fermion evolves independently from the other, feeling a constant force pulling it towards the reflective barrier. Hence, in between two scatterings, the left fermion obeys the equation of motion $\dot{k}=\chi\, ,\dot{x}=v(k) $ (we refer to the figure below for notation)

\begin{figure}[h!]
\includegraphics[width=0.3\textwidth]{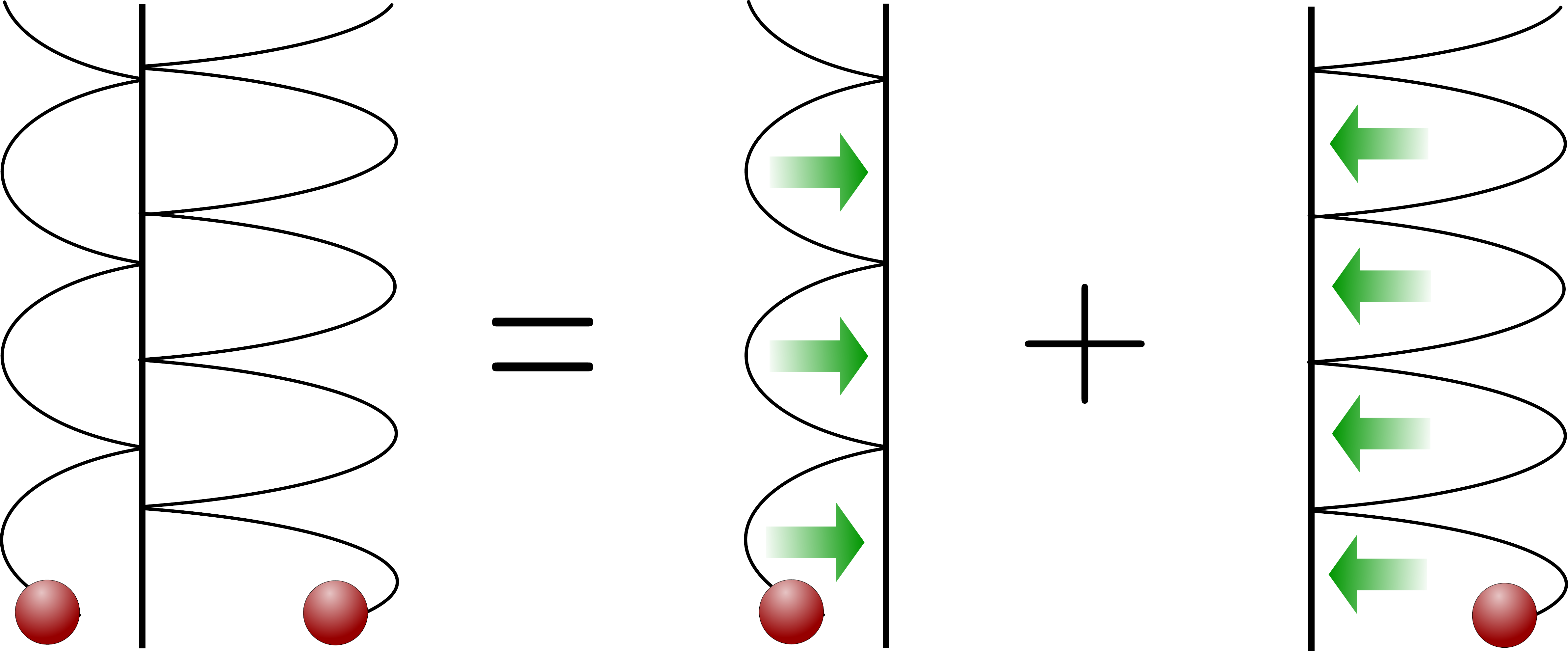}
\end{figure}

The period of the oscillation is readily computed noticing that, right after the reflection, the momentum of the fermion changes sign $k\to -k$ and the oscillation period $t_{osc}(k)$ is the time needed for the force to bring the momentum back to $k$, i.e. $k=-k+\chi t_{osc}(k)$. We now consider the probability that the leftmost fermion is not transmitted after $n$ scatterings or, equivalently, the probability of being reflected $R(k)^n=(1-T(k))^n\simeq \exp[-nT(k)]\simeq \exp[-T(k) t/t_{osc}(k)]$. Using the oscillation period and asking that both fermions are not transmitted until time $t$, one gets Eq. \eqref{eq_M_lifetime} for the time evolution of the trapped probability, i.e.
\be\label{S_eq_M_lifetime}
P_t(k,q)=\exp\left[-t\frac{\chi}{2}\big(|k|^{-1} T(k)+|q|^{-1}T(q)\big)\right]P_{t=0}(k,q)\, .
\ee

The initial probability $P_{t=0}(k,q)$ depends on the details of the scattering, but surprisingly it can be computed in terms of geometrical considerations without solving the equation of motion.

We now consider the probability of forming a metastable state by shooting a mesonic wavepacket at the defect. For the sake of simplicity, we assume the meson has a well defined energy $E$ and total momentum $K$. Furthermore, we approximate the capture time to be negligible (i.e. all the mesons of the wavepacket are captured within a time window much smaller than the decay time) and set $t=0$ as the scattering time.
Also within the assumption that the meson is captured only after a single transmission event, there are several possible processes where some reflections take place before the desired transmission shown below.

\begin{figure}[h!]
\includegraphics[width=0.5\textwidth]{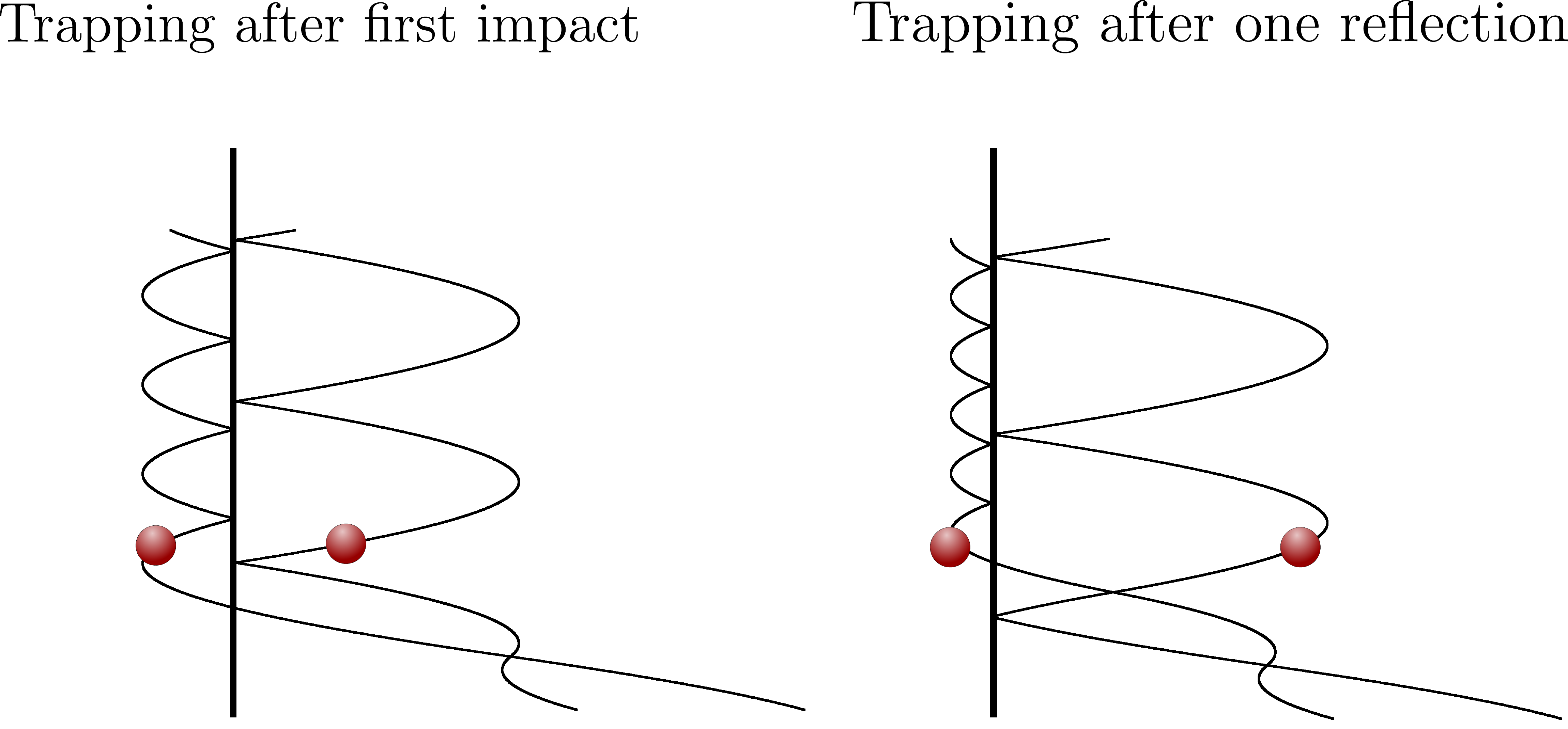}
\end{figure}

Each of these processes leads a different pair $\{(k_\ell,q_\ell)\}_{\ell=1,2,...}$ of momenta for the trapped meson.
The momenta $k_{\ell}$ and $q_{\ell}$ are not independent, but they must satisfy the constraint $E=\epsilon(k_{\ell})+\epsilon(q_{\ell})$: we prove it in the case of $\ell=1$ with the aid of the picture below, the argument is easily extended to the case of generic $\ell$.
\begin{figure}[h!]
\includegraphics[width=0.3\textwidth]{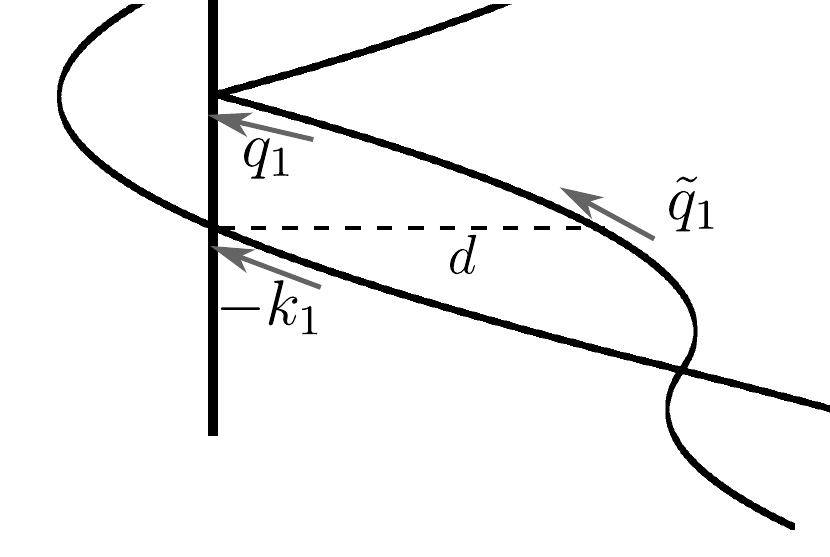}
\end{figure}
Let us focus on the first transmission event: in our notation, the impact of the first transmitted fermion happens at momentum $-k_{1}$. Meanwhile, the companion fermion is placed at a distance $d$ and with momentum $\tilde{q}_{1}$.
The total energy of the meson is thus $E=\epsilon(-k_{1})+\epsilon(\tilde{q}_{1})+\chi d$, for simplicity we can use the energy parity $\epsilon(-k_{1})=\epsilon(k_{1})$.
After the first fermion gets transmitted, the other will move independently with momentum $q(t)$ and distance from the defect $d(t)$ and obeying the equation of motion $\dot{q}=-\chi$ $\dot{d}=v(q(t))$. Of course, $E'=\epsilon(q(t))+\chi d(t)$ is a conserved quantity. By comparison, with the energy of the meson, we have $E'=E-\epsilon(\tilde{q}_{1})$. On the other hand, $q^{(1)}$ is the momentum at the moment of impact and can be found by asking $d(t)=0$, hence $E=\epsilon(q_{1})+\epsilon(k_{1})$.
The same argument can be generalized to arbitrary $\ell$ by noticing that the scattering with the defect conserves the total energy of the meson.

Energy conservation allows one to find $q_{\ell}$ if $k_{\ell}$ is known. As a next step, we build a recursive set of equations that fixes $k_\ell$ from the knowledge of the moment at first impact $k_1$, together with the total energy and momentum of the fermion.
With the help of the figure below, let us consider $-k_1$ the momentum at first impact and, as before, let $\tilde{q}_1$ be the momentum of the companion. Right before the impact, the total momentum $K=-k_1+q_1$ then the scattering fermion gets reflected and the total momentum becomes $K\to K'= k_1+\tilde{q}_1$. 
Our next task is finding the momenta $\bar{k},\bar{q}$ at the moment of the scattering among the two fermions. This is easily found by imposing that the total energy of of the meson is purely kinetic and using the conservation of $K'=\bar{k}+\bar{q}$
\be
E=\epsilon(\bar{k})+\epsilon(k_1+\tilde{q}_1-\bar{k}).
\ee

\begin{figure}[h!]
\includegraphics[width=0.4\textwidth]{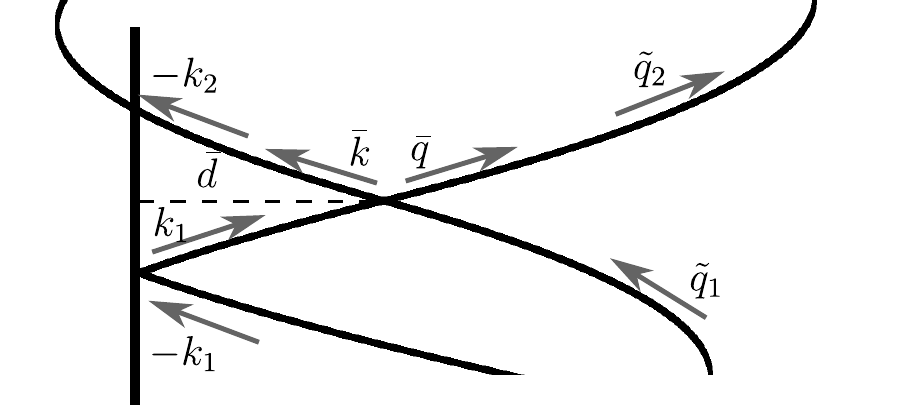}
\end{figure}

Lastly, from $\bar{k}$ and energy conservation one finds $-k_2$. Let be $\bar{d}$ the distance between the defect and the position of the scattering among the two fermions. With the current convention on the momenta, one has $\epsilon(\bar{q})=\epsilon(k_1)+\chi d$, but also $\epsilon(\bar{k})=\epsilon(k_2)+\chi d$, whose comparison gives the simple equation $\epsilon(\bar{k})+\epsilon(\bar{q})=\epsilon(k_1)+\epsilon(k_2)$.
In summary, and moving to the general case $k_\ell$, the recursive relation $k_\ell\to k_{\ell+1}$ is found as the solution of
\be
\begin{cases}
E=\epsilon(\bar{k}_\ell)+\epsilon(k_\ell+\tilde{q}_\ell-\bar{k}_\ell),\\
\epsilon(\bar{k}_\ell)+\epsilon(k_\ell+\tilde{q}_\ell-\bar{k})=\epsilon(k_\ell)+\epsilon(k_{\ell+1}),\\
-k_{\ell+1}+\tilde{q}_{\ell+1}=k_\ell+\tilde{q}_\ell.
\end{cases}
\ee
Since from momentum conservation one has $\tilde{q}_1=K-k_1$, the full sequence $(k_\ell,q_\ell)$ is entirely determined by $k_1$.

Let $p(k_1)$ being the probability that the meson hits the impurity with a fermion of momentum $-k_1$. Hence, it will create a trapped meson $(k_1,q_1)$ with probability $p(k_1)T(-k_1)$. The pair $(k_2,q_2)$ is instead created if the first scattering is reflective and then transmissive, hence it will be excited with probability $p(k_1)(1-T(-k_1)) T(-k_2)$ and so on so forth.
Eventually, the probability $P_t(E,K)$ that a meson with energy $E$ and momentum $K$ gives a metastable bound state at time $t$ is
\be\label{eq_P_classic}
P_t(E,K)=\int \dd k\, p(k_1) \sum_{\ell=0}^\infty T(-k_\ell)\Bigg\{\prod_{i=1}^{\ell-1}[1-T(-k_i)]\exp\left[-\frac{\chi t}{2}\left(|k_\ell|^{-1}T(k_\ell)+|q_\ell|^{-1}T(q_\ell)\right)\right]\Bigg\}_{E=\epsilon(k_\ell)+\epsilon(q_\ell)}\, .
\ee
The last ingredient is determining $p(k_1)$ that in general depends on the fine details of the initial state and the whole time evolution. However, there is at least one case where $p(k_1)$ can be easily obtain, i.e. in the approximation that the size of the wavepacket is much larger than the typical size of the meson. If it is the case, the impact of the fermion will randomly happen at a given position in its trajectory. In the figure below (left), the position of the defect can be arbitrarily moved within a maximum interval of length $D$.
Notice that $D$ is nothing else than the distance traveled by the meson within his breathing period, which we can compute with the help of the right figure.

\begin{figure}[h!]
\includegraphics[width=0.5\textwidth]{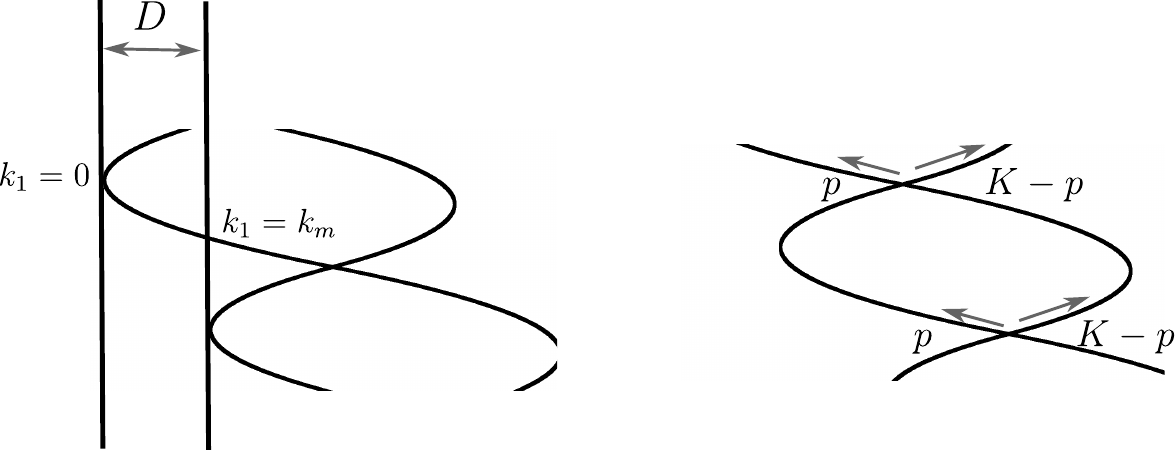}
\end{figure}
Let $p$ and $K-p$ the momenta of the two fermions when they scatter. Hence, by conservation of energy $p$ is fixed solving $E=\epsilon(p)+\epsilon(K-p)$. Let us follow the trajectory of the fermion with initial momentum $p$: after an oscillation period and right before scattering with the companion, its momentum will be $K-p$. Let us now focus on the displacement: from the equation of motion, we have $D=\left|\int_0^\tau \dd t v(p(t))\right|$ with $\tau$ the oscillation period and $p(t)$ the time-evolving momentum. By using $\dot{p}=\chi $ and $v(k)=\partial_k\epsilon(k)$, we can easily compute $D$ as $ D=\chi^{-1}|\epsilon(K-p)-\epsilon(p)|$.
Finally, we can compute $k_m$ from energy conservation $\epsilon(k_m)=\epsilon(0)+\chi D$. Putting the pieces together, $k_m$ is defined by solving
\be
\begin{cases}
\epsilon(k_m)=\epsilon(0)+|\epsilon(K-p)-\epsilon(p)|,\\
E=\epsilon(p)+\epsilon(K-p).
\end{cases}
\ee

Within the allowed window of momenta, we are considering a flat average over the position of the meson at the moment of impact.
When changing coordinates to the momentum space, one needs to consider the proper Jacobian: this is easily done from the equation of motion $|\dot{k}|=\chi$ and $|\dot{x}|=|v(k)|$, hence $\dd k_1=\dd x |v(-k_1)| \chi^{-1}$.
Thus, one finds
\be
p(-k_1)\propto|v(-k_1)|\theta(|k_m|-|k|)\theta(k_1\text{sign}(v(k_1)))\, ,
\ee
where $\theta$ is the Heaviside theta function and the proportionality constant is fixed by imposing that $p(-k_1)$ is normalized to unity.

\section{The tensor network simulations}
\label{S_sec_Tensor}

In this section we discuss the tensor network simulations and details for Fig. \ref{fig_tensor_networks}. However, first we discuss how the initial mesonic wavepacket can be prepared and controlled.
\ \\ \ \\
\textbf{The wavepacket initialization: the meson creation operator ---} 
Here we discuss how to create two fermions in the transverse Ising chain via spin operators. It will be later employed for the creation of a moving meson as the initial state in the tensor network numerical simulation. We start by considering the transverse field Ising model
\be\label{eq_Ising}
H=-\sum_j \frac{1}{2}\left[\sigma_j^z\sigma_{j+1}^z+h_x \sigma_j^x\right],
\ee
The system can be mapped into the free fermion representation with a Jordan Wigner transformation
\be
d_j=\sum_{i=-\infty}^{j-1}\sigma_i^x\sigma_j^-
\ee
with $\sigma_j^\pm=(\sigma_j^z\pm i\sigma_j^y)/2$.
Then one defines the modes in the Fourier space as
\be
\begin{pmatrix}
d_j \\ d_j^\dagger\end{pmatrix}=\int_{-\pi}^\pi \frac{\dd k}{2\pi} e^{ik j}\begin{pmatrix}\cos\theta_k&&i \sin\theta_k \\ i\sin\theta_k && \cos\theta_k \end{pmatrix}\begin{pmatrix}\gamma(k) \\ \gamma^\dagger(-k) \end{pmatrix},
\ee
where $\theta_{k}=\frac{1}{2i}\log\left[(h_x-e^{ik})/(1+h_x^2-2h_x\cos(k))\right]$. With this choice, the Ising Hamiltonian is diagonal in the mode operators $\gamma(k)$ and the ground state is identified with the vacuum $\gamma(k)|0\rangle=0$.
As a next step, we would like to create a pair of fermions on top of the vacuum. To this end, let us consider an operator defined in the following form
\be
O_j=\sum_{\ell\ge 0} \sigma^+_{j}\prod_{i=j}^{j+\ell-1}\sigma_i^x \sigma^+_{j+\ell} F(\ell)=\sum_{\ell\ge 0} d^\dagger_{j} d^\dagger_{j+\ell} F(\ell),
\ee
with a fast decaying, e.g. exponential, space-dependent function $F(\ell)$. 
From the $O_j$ operator, we create $O(P,j_0)$ as
\be
\label{eq.gaussianstring}
O(P,j_0)=\sum_j e^{-(j-j_0)^2/\sigma^2}e^{i Pj}( O_j+\mathcal{A} ).
\ee
The idea is that $O(P,j_0)$ tries to create a wavepacket in motion, centered around $j_0$ and with momentum $P$. $\mathcal{A}$ is a constant inserted for imposing the normal ordering with respect to the $\gamma$ operators and is determined below.

When we express $O_j$ in terms of modes, it will contain operators in the form $\gamma^\dagger \gamma^\dagger$ (which create two fermions), but also $\gamma^\dagger\gamma$, $\gamma \gamma^\dagger$ and $\gamma\gamma$.
We are interested in the action of $O(P,j_0)$ on the vacuum and if we fix the constant $\mathcal{A}$ as
\be
\mathcal{A}=-\Bigg(\sum_{\ell>0} F(\ell) \int \frac{\dd q}{2\pi} e^{-iq\ell} \left(i\sin\theta_q \cos\theta_q \right)\Bigg),
\ee
we obtain a two-fermions state
\be
O(P,j_0)|0\rangle=\sum_j e^{-(j-j_0)^2/\sigma^2} e^{iP j}
\Bigg(\sum_{\ell\ge0} F(\ell) \int \frac{\dd k}{2\pi}\int \frac{\dd q}{2\pi} e^{-ikj}e^{-iq(j+\ell)} \cos\theta_k\cos\theta_q \gamma^\dagger(k)\gamma^\dagger(q)\Bigg)|0\rangle.
\ee

We now first sum over $j$ and define
\be
W(P-k-q)=\sum_j e^{-(j-j_0)^2/\sigma^2} e^{i(P-k-q) j},
\ee
which becomes very peaked in the momentum space $\lim_{\sigma\to \infty} W(P-k-q)=2\pi \delta(P-k-q)$ for large $\sigma$.
We further define
\be
\tilde{F}(q)=\sum_{\ell\ge 0} F(\ell)  e^{-i q\ell},
\ee
leading to the following compact expression 
\be
O(P,j_0)|0\rangle=
\Bigg( \int \frac{\dd k}{2\pi}\int \frac{\dd q}{2\pi} W(P-k-q)\tilde{F}(q) \cos\theta_k\cos\theta_q \gamma^\dagger(k)\gamma^\dagger(q)\Bigg)|0\rangle.
\ee
Lastly, we use the asymmetry of the fermions to rewrite the last expression as

\be
O(P,j_0)|0\rangle=
\Bigg( \int \frac{\dd k}{2\pi}\int \frac{\dd q}{2\pi} W(P-k-q)\frac{\tilde{F}(q)-\tilde{F}(k) }{2}\cos\theta_k\cos\theta_q \gamma^\dagger(k)\gamma^\dagger(q)\Bigg)|0\rangle.
\ee
Clearly, $\left|W(P-k-q)\frac{\tilde{F}(q)-\tilde{F}(k) }{2}\cos\theta_k\cos\theta_q \right|^2$ is the semiclassical probability of the wavepacket.
Tuning $\sigma$ and the function $F$ we can change the wavepacket. So far we kept $F$ arbitrary but we require $F$ to be fast decaying in $\ell$ for both computational reasons and to get a more localized wavepacket.
$F$ must be tuned if one wants to act on the probability distribution of the difference in the momentum of the two fermions (while $W$ acts on the total momentum).
A convenient choice is choosing $F$ as a Kronecker delta
\be
F(\ell)=\delta_{\ell,\ell_0}
\ee
for a certain $\ell_0$. Note, the state $|\Psi\rangle=O(P,j_0)|0\rangle$ is not normalized. It must be renormalized before running the simulation. After this, the state $|\Psi\rangle$ will exactly contain two fermions, i.e. one meson.
\ \\ \ \\ \ \\
\textbf{Numerical details on the Tensor network calculation ---} The tensor network simulation of the dynamics is implemented via the Python library TeNPy as detailed in Ref.~\cite{hauschild2018efficient}. The strong confinement in our mode significantly suppresses the spreading of correlations throughout the whole system, hence, permitting an efficient tensor network simulation with a low bond dimension for a long time.

We first use the Density Matrix Renormalization Group (DMRG) algorithm to prepare the system of length $L=240$ with open boundary in its groundstate of the Hamiltonian
\be
H_\text{Ising}=-\sum_i \sigma_i^z\sigma_{i+1}^z - h_x\sum_i \sigma_i^x - (d-h_x) \sigma_0^x- h_z\sum_i \sigma_i^z,
\ee
where the defect located on site $0$. 
The transverse field is chosen as $h_x=0.3$, and a non-zero but small longitudinal field $h_z=0.01$ is used to break the two-fold groundstate degeneracy. As the groundstate is approximately a simple ferromagnetic state without long-range correlation, a low bond dimension $\chi=20$ is sufficient. 

Now we want to create the initial meson wavepacket with a non-zero velocity such that it can move towards the impurity. As introduced in the last section, we construct a string operator $O\left(P, j_{0}\right)$ according to Eq.~\eqref{eq.gaussianstring},
\begin{equation}
\label{eq.wavepacket}
    O\left(P, j_{0}\right)=\sum_{j} e^{-\left(j-j_{0}\right)^{2} / \sigma^{2}} e^{i P j}\left(O_{j}+\mathcal{A}\right),
\end{equation}
with 
\be
\label{eq.operator}
O_{j}=\sum_{\ell \geq 0} \sigma_{j}^{+} \prod_{i=j}^{j+\ell-1} \sigma_{i}^{x} \sigma_{j+\ell}^{+} F(\ell),\ \ \  \mathcal{A}=-\Bigg(\sum_{\ell>0} F(\ell) \int \frac{\dd q}{2\pi} e^{-iq\ell} \left(i\sin\theta_q \cos\theta_q \right)\Bigg),
\ee
and $F(\ell)=\delta_{l,1}$.
 Numerically we choose $j_0=10$, $P=\pi/2$, $\sigma=\sqrt{4L/\pi}$ and the summation over $j$ to be limited within $[j_0-10,j_0+10]$.
 Acting the operator $O\left(P, j_{0}\right)$ on the groundstate creates a wavepacket with a total momentum $P$. However, the created meson is a superposition of excitations of different energies, and therefore, each of them has a different velocity. 
\begin{figure}[h!]
\includegraphics[width=0.3\textwidth]{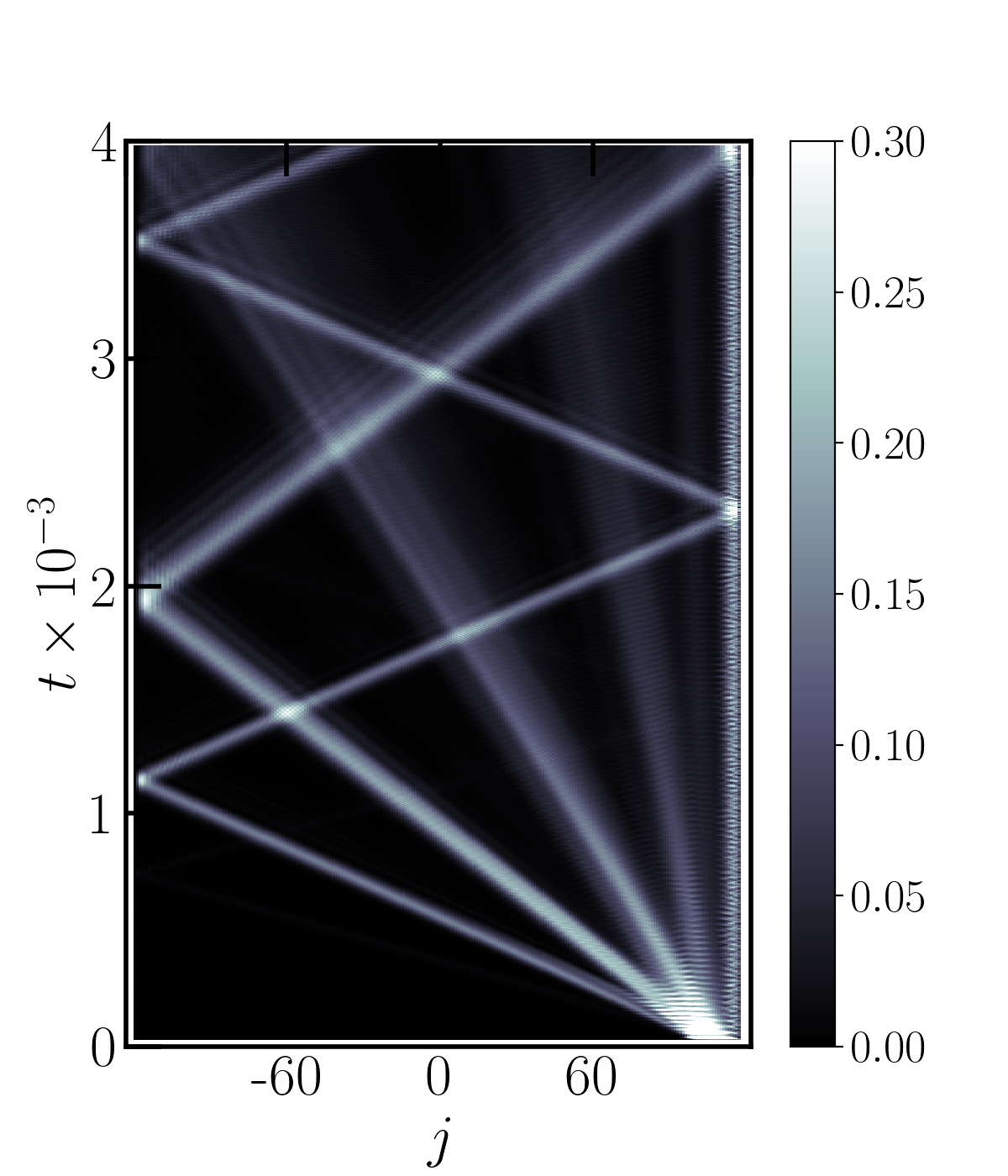}
\includegraphics[width=0.275\textwidth]{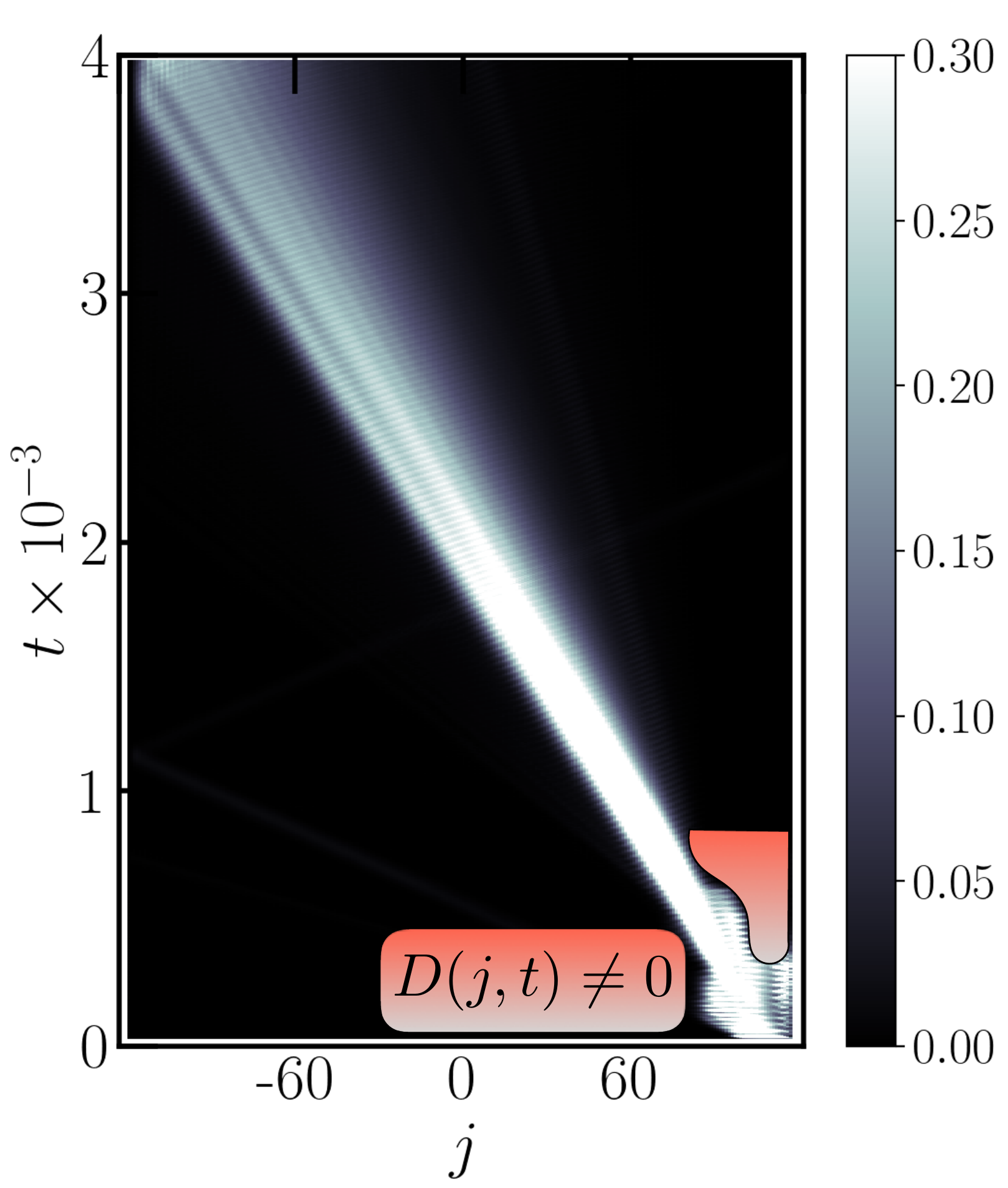}
\includegraphics[width=0.3\textwidth]{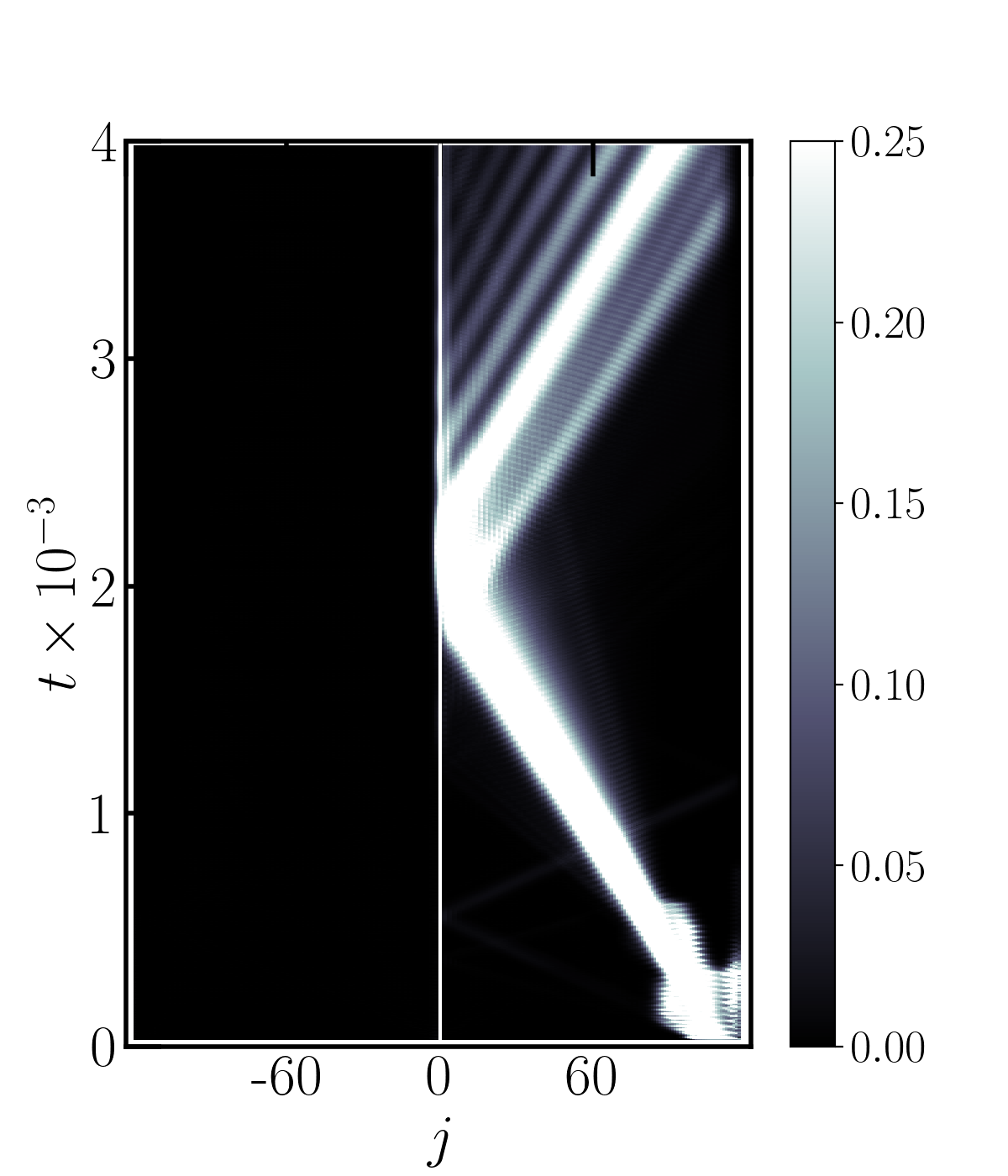}
\caption{Dynamics of the connected correlation $\langle \sigma^z_i\sigma^z_{i+1}\rangle$ calculated via TEBD algorithm. Left: no dissipation and no impurity. Middle: \hz{No impurity $d=h_x$}. Dissipation is included such that only one wavepacket survives. Right: Both dissipation and impurity \hz{d=0.1} are present and the metastable state is clearly visible. For all three plots, we use $h_x=0.3,h_z=0.12,L=240$.}
\label{fig:bareTEBD}
\end{figure}

The time evolution of the dynamics is achieved by the Time Evolving Block Decimation (TEBD) algorithm with a low bond dimension $\chi=20$. We use a time step $dt=1$ and the fourth order Suzuki-Trotter decomposition for the time evolution. The connected part of the correlation $\langle \sigma^z_i\sigma^z_{i+1}\rangle$ is employed to trace the position of the meson. As shown in the left panel of Fig.~\ref{fig:bareTEBD}, the wavepacket quickly spreads with distinct velocities, corresponding to different excitations of the system. 

In the absence of the impurity, the wavepackets with the faster velocities bounce at the boundary and reflect back, interfering with the slowly moving wavepackets at later times. Although not shown here, such interference happens more severely when the impurity is present. Consequently, the trapped meson is not clearly visible and its lifetime is difficult to analyze.
To address this, we further additionally introduce the non-Hermitian Hamiltonian to induce week dissipation of the form
\be
H_\text{dis}(t)=i\sum_j D(j,t) \left(\sigma_j^z\sigma_{j+1}^z + h_x \sigma_j^x\right),
\ee
where $D(j,t)$ is a time-dependent function with a positive amplitude (the maximum value is around 0.01) that smoothly decays in space. For a fixed time $t_0<750$, $D(j,t_0)$ is non-zero in regions where undesired meson components move through and get dissipated, permitting us to select the meson wavepacket with an approximately constant velocity. As shown in the middle panel of Fig.~\ref{fig:bareTEBD}, we choose a profile $D(j,t)$ nonzero in the red regions such that only one wavepacket survives and the amplitude of the correlation function also gets amplified due to the normalization of the wavefunction. At later times, we also use a non-zero $D(j,t)$ at the boundary to reduce the finite size effect. In the end, in the right panel, we introduce the impurity  $d=0.1$ same as Fig. \ref{fig_tensor_networks} at the middle of the system and plot the dynamics where a long-lived metastable state can be clearly identified.
\begin{figure}[h!]
\label{fig:entdynamics}
\includegraphics[width=0.5\textwidth]{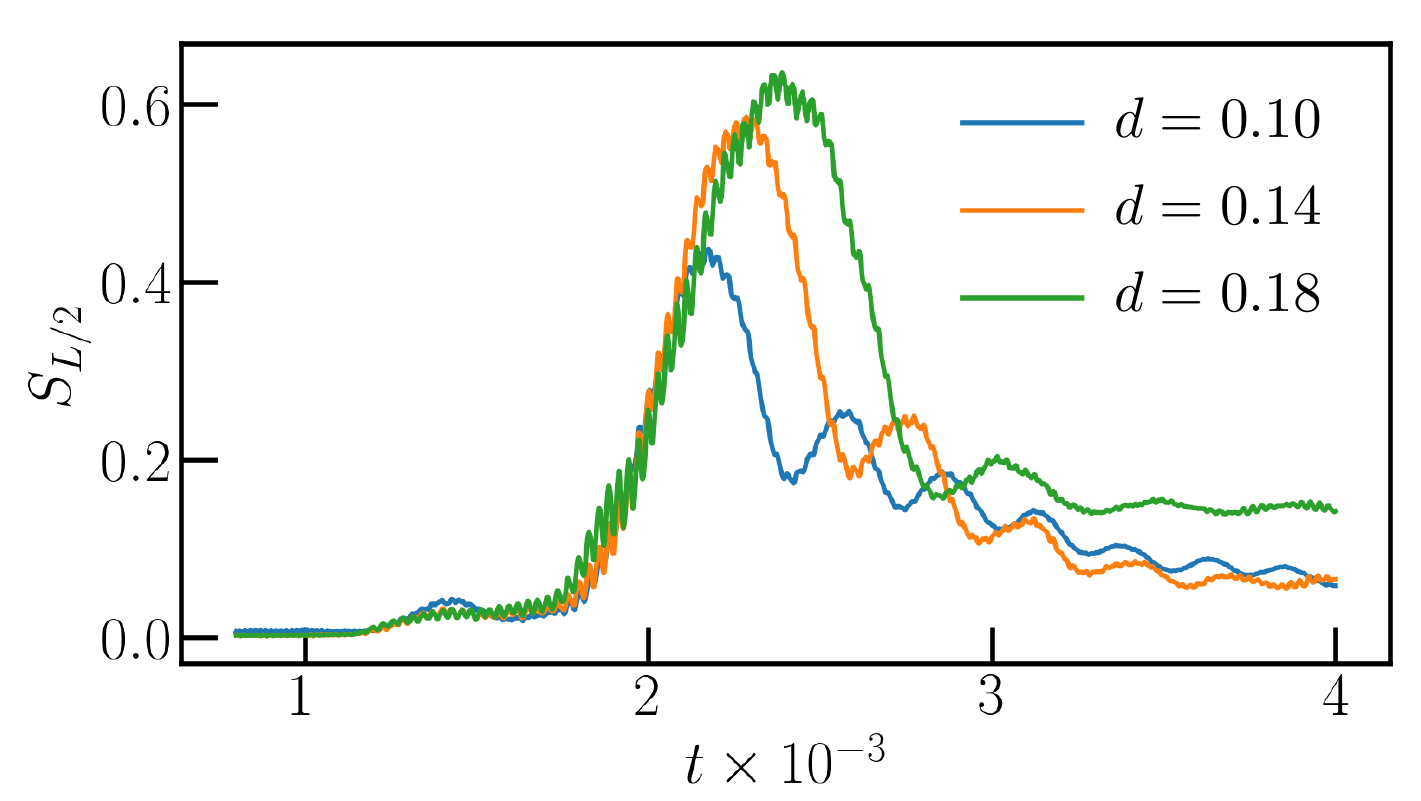}
\caption{Dynamics of the half-system entanglement entropy.}
\end{figure}

We further provide the dynamics of the half-system entanglement entropy defined as
\begin{equation}
    S_{L/2}(A)=-\operatorname{tr} \hat{\rho}_{L/2} \log \hat{\rho}_{L/2},
\end{equation}
where $\hat{\rho}_{L/2}$ denotes the reduced density matrix of half of the system. As the impurity locates at the center of the chain, before the meson-impurity scattering happens, there is almost no entanglement established between two halves of the system. Entanglement entropy suddenly increases around $t\sim 2\times 10^3$ where one kink tunnels through the impurity and becomes entangled with the other kink reflected back. It drops down when the transmitted and reflected particles eventually leave the impurity. Overall, the entanglement of the whole system remains at low values permitting the efficient long-time simulation of the dynamics with a low bond dimension.
\ \\ \ \\ \ \\
\textbf{Details on Fig. \ref{fig_tensor_networks} ---} Here we give a brief summary of parameters used for Fig. \ref{fig_tensor_networks}. For the initial state generation, we use $F(l)=\delta_{l,1}, j_0=10,$ $P=\pi/2$ and $\sigma=\sqrt{4L/\pi}$ to determine string operator in Eq.~\ref{eq.operator}. For the time evolution, we use bond dimension $\chi=20$, $dt=1$ and the fourth order Suzuki-Trotter decomposition. The Hamiltonian parameters are $h_x=0.3,h_z=0.12$ and $L=240$. The defect size is $d=0.1$ in the left panel.

\end{document}